\documentclass[12pt]{iopart}
\usepackage{iopams}

\usepackage{epsfig}
\usepackage{epstopdf}
\usepackage{epsf}
\usepackage{subfigure}
\usepackage{graphicx}
\usepackage{bm}
\usepackage{float}
\usepackage{sidecap}
\usepackage{hyperref}
\usepackage{ulem}
\usepackage{xcolor}
\usepackage{wrapfig}
\usepackage{ulem}

\def \bfx {{\bf x}}
\def \bfu {{\bf u}}

\def \bfb {{\bf b}}
\def \bfa {{\bf a}}
\def \bfk {{\bf k}}
\def \bfl {{\bf l}}

\def \bfj {{\bf j}}

\def \bomega {{\bm \omega}}

\def \f {{\bm f}}

\def \lap {\nabla^2}

\begin{document}
\title[DNS of 3DMHD turbulence with random power-law forcing]{Direct Numerical
Simulations of Three-dimensional Magnetohydrodynamic Turbulence
with Random, Power-law Forcing}
\author{Ganapati Sahoo~$^1$, Nadia Bihari Padhan~$^2$, Abhik Basu~$^3$, and Rahul Pandit~$^{2,4}$}
\address{
$^1$ Department of Mathematics and Statistics, University of Helsinki\\
$^2$ Centre for Condensed Matter Theory, Department of Physics,
Indian Institute of Science,
Bangalore 560012, India. \\
$^3$ Theoretical Condensed Matter Physics Division,
Saha Institute of Nuclear Physics, Calcutta, India.\\
$^4$ Also at: Jawaharlal Nehru Centre for
Advanced Scientific Research, Jakkur,
Bangalore, India

}
\ead{\mailto{ganapati.sahoo@gmail.com};
\mailto{nadia@iisc.ac.in};
\mailto{abhik123@gmail.com};
\mailto{rahul@iisc.ac.in}}

\begin{abstract} We present pseudospectral direct-numerical-simulation (DNS)
studies of the three-dimensional magnetohydrodynamic (MHD) equations
(3DRFMHD) with a stochastic force that has zero mean and a variance
$\sim k^{-3}$, where $k$ is the wavenumber, because 3DRFMHD is used in field-theoretic
studies of the scaling of energy spectra in MHD turbulence. We obtain velocity and
magnetic-field spectra and structure functions and, from these, the multiscaling
exponent ratios $\zeta_p/\zeta_3$, by using the extended self similarity
(ESS) procedure. These exponent ratios lie within error bars of their
counterparts for conventional three-dimensional MHD turbulence (3DMHD).
We then carry out a systematic comparison of the statistical properties of
3DMHD and 3DRFMHD turbulence by examining various probability
distribution functions (PDFs), joint PDFs, and isosurfaces of
of, e.g., the moduli of the vorticity and the current density for
three magnetic Prandtl numbers ${\rm Pr_M} = 0.1,~1$, and $10$.  
\end{abstract}

\maketitle

\section{Introduction \label{sec:intro}}

The elucidation of the statistical properties of turbulence continues to
fascinate scientists~\cite{book-frisch,pramana09} for it poses very
challenging problems, experimental, numerical, and theoretical, that lie at
the interfaces between nonequilibrium statistical mechanics, the nonlinear
dynamics of extended systems, and fluid dynamics. This fascination is not
restricted to fluid turbulence, for which we use the Navier-Stokes equation,
but it also extends to other forms of turbulence, such as that in conducting
fluids, for which we use the equations of magnetohydrodynamics
(MHD)~\cite{book-arnab,book-vkrishan,book-rudiger,book-goedbloed,book-biskamp,mkvrev04} and
on which we concentrate here. The flows of such conducting fluids is often
turbulent in a variety of physical settings, including liquid-metal flows in
terrestrial laboratories and planetary
interiors~\cite{dormy08,roberts00,fauve01,riga00,karl01,peffley00,pinton07},
in stars such as the
sun~\cite{book-arnab,book-vkrishan,book-goedbloed,book-biskamp}, in
the solar wind~\cite{salem09,podesta09}, and in the interstellar
medium~\cite{book-vkrishan,book-rudiger,book-goedbloed}.

Large kinetic and magnetic Reynolds numbers, ${\rm Re}=UL/\nu$ and ${\rm
Re_M}=UL/\eta$, respectively, lead to turbulent motion of the conducting fluid;
here $L$ and $U$ are typical length and velocity scales in the flow, $\nu$ is
the kinematic viscosity, and $\eta$ is the magnetic diffusivity. The
statistical characterization of MHD turbulence is harder than its
fluid-turbulence counterpart because (a) we must control  ${\rm Re}$ and ${\rm
Re_M}$ and (b) we must obtain the statistical properties of both velocity and
magnetic fields. Furthermore, $\nu$ and $\eta$ can differ by several orders of
magnitude, so the magnetic Prandtl number ${\rm Pr_M}\equiv {\rm Re_M}/{\rm Re}
= \nu/\eta$ can vary over a large range: ${\rm Pr_M}\simeq 10^{-5}$ in the
liquid-sodium system~\cite{riga00,karl01}; ${\rm Pr_M}\simeq 10^{-2}$ at the
base of the Sun's convection zone~\cite{schekochihin04prl}; and ${\rm
Pr_M}\simeq 10^{14}$ in the interstellar medium~\cite{elmegreen04,ponty05}.
Hence, the fluid-dissipative and magnetic-resistive scales, $\ell_d^u$ and
$\ell_d^b$, respectively, can be well separated in MHD turbulence:
$\ell_d^u\sim \nu^{3/4}$ and $\ell_d^b\sim \eta^{3/4}$ at the level of
Kolmogorov 1941 (K41) phenomenology~\cite{book-frisch,k41}). Careful studies of
the statistical properties of MHD turbulence must resolve both these
dissipative scales, a daunting task at large ${\rm Re}$, especially if ${\rm
Pr_M}$ is significantly different from unity. Most direct numerical simulations
(DNS) of MHD
turbulence~\cite{kalelkar04,mininni06,mason08,baerenzung08,mininni09,biskamp00,alex13,bisk}
have been restricted to ${\rm Pr_M}\simeq 1$; but some DNSs have now started
moving away from the ${\rm Pr_M}\simeq 1$
regime~\cite{ponty07,brandenburg09,sahoomhd,sahoophd}.

We investigate the properties of incompressible MHD turbulence in three
dimensions (3D), in the absence of a mean magnetic field and with a specific
type of power-law, random forcing, which we define precisely below; henceforth,
we refer to this type of MHD as 3DRFMHD, as opposed to its conventionally
forced counterpart, which we refer to as 3DMHD. The motivation for our study
arises from the use of such random forcing for field-theoretical studies,
initially of fluid
turbulence~\cite{dedominicis79,yakhot86,jkb88,book-adzhemyan,busa97,Olla} and
subsequently of MHD
turbulence~\cite{mkvrev04,fournier82,adzhemyan85,hnatic95,hnatich01,kim04,kim99,zhou88,zhou}.
It is useful, therefore, to examine the statistical properties of 3DRFMHD
turbulence by a careful, direct numerical simulation (DNS). Such DNS studies
have been carried out, with this type of power-law, random forcing, for the
one-dimensional Burgers~\cite{chekhlov95,hayot96,mitra05}, three-dimensional
Navier-Stokes~\cite{sain98,biferale04}, and two-dimensional Navier-Stokes
equations~\cite{mazzino09} but not, so far, for 3DRFMHD turbulence. One of the
principal goals of our study is to investigate whether the inertial-range
scaling properties of energy spectra in 3DRFMHD are the same as those of 3DMHD.
We also present data for 3DRFMHD for (a) the temporal evolution of quantities
such as the energy, energy-dissipation rates, Reynolds numbers, and important
length scales, (b) energy, dissipation-rate, and effective-pressure spectra,
(c) probability distribution functions (PDFs) of components of various fields,
energy dissipation rates, and cosines of angles between various vectors, (d)
velocity and magnetic-field structure functions, which can be used to
characterize inertial-range intermittency, (e) isosurfaces of
the moduli of the vorticity and the current, the energy-dissipation rates, and
the effective pressure, and (f) joint PDFs that characterize the statistical
correlation between various quantities. Field-theoretical studies have not, so
far, computed as many quantities as we calculate for 3DRFMHD; thus, a
comparison of our results here with those available from DNS studies of
3DMHD~\cite{sahoomhd,sahoophd} provides a guide to which statistical properties
of 3DMHD might emerge from analytical but, perforce, approximate studies of
3DRFMHD. Our study examines the magnetic-Prandtl-number dependence of 3DRFMHD,
by carrying out DNSs at ${\rm Pr_M}=0.1,~1,$ and $10$.  We follow the 3DMHD
investigations of Ref.~\cite{sahoomhd,sahoophd}, so we do not include (a)
compressible MHD turbulence, (b) a mean magnetic field (as, e.g., in
Ref.~\cite{goldreich95}) and (c) Lagrangian properties (see, e.g.,
Ref.~\cite{homann07}). 

Such a comprehensive study of 3DRFMHD turbulence has not been attempted
hitherto.  Given our error bars, we find that there is agreement between our
3DRFMHD exponent ratios with those of 3DMHD~\cite{sahoomhd,sahoophd}.  The
${\rm Pr_M}$ dependence of these exponent ratios is also similar for 3DRFMHD
and 3DMHD turbulence. This is of importance in deciding whether 3DRFMHD can be
used for obtaining universal statistical properties, such as exponent ratios,
by employing field-theoretic renormalization-group methods.  Although these
exponent ratios for 3DRFMHD and 3DMHD seem to agree, some PDFs are different in
detail and the isosurfaces mentioned above are qualitatively different.
Furthermore, strictly speaking (see below), 3DMHD and 3DRFMHD turbulence cannot
be in the same universality class because of logarithmic corrections to, say,
the energy spectrum because of the power law in the random forcing we employ
(for a discussion in the analogous randomly forced versions of the 3D
Navier-Stokes equation and the 1D Burgers equation see
Refs.~\cite{mitra05,sain98}). 

We organize the remaining part of this paper as follows:  In
Sec.~\ref{sec:equations} we describe the 3DRFMHD equations, the forcing method
and the numerical scheme we use, and the statistical measures we employ to
characterize 3DRFMHD turbulence. In Sec.~\ref{sec:results} we present our
results in several subsections: Subsection~\ref{sec:tseries} is devoted to the
temporal evolution of quantities such as the energy and energy-dissipation
rates, Reynolds number, and important length scales; in
Subsec.~\ref{sec:spectra} we present energy, dissipation-rate, and
effective-pressure spectra; in Subsec.~\ref{sec:pdfs} we show probability
distribution functions (PDFs) of quantities such as components of fields,
energy-dissipation rates, and cosines of angles between various vectors; in
Subsec.~\ref{sec:stfn} we present  velocity and magnetic-field structure
functions and explain how they can be used to characterize inertial-range
intermittency; in Subsec.~\ref{sec:isosurface} we display the isosurfaces
mentioned above; in Subsec.~\ref{sec:jpdfs} we present joint PDFs that quantify
the statistical correlation between various quantities.
Section~\ref{sec:conclusions} contains a discussion of our results.

\section{Equations, Numerical Methods, and Statistical Measures \label{sec:equations}}

The 3DRFMHD model is defined by the following set of equations:
\begin{eqnarray}
\frac{\partial\bfu}{\partial
t}+(\bfu\cdot\nabla)\bfu=\nu\lap\bfu
-\nabla\bar{p}+(\bfb\cdot\nabla)\bfb+\f_u, \\ \label{magone}
\frac{\partial\bfb}{\partial t}+(\bfu\cdot\nabla)\bfb
=(\bfb\cdot\nabla)\bfu+\eta\lap\bfb+\f_b ; \label{eq:magtwo}
\end{eqnarray}
$\bfu$, $\bfb$, $\bomega=\nabla\times\bfu$
and $\bfj=\nabla\times\bfb$ are, respectively, the velocity
field, the magnetic field, the vorticity, and the current density
at the point $\bf{x}$ and time $t$; $\nu$ and $\eta$ are the
kinematic viscosity and the magnetic diffusivity, respectively,
and the effective pressure is $\bar{p}=p+(b^2/8\pi)$, where $p$
is the pressure. We enforce the incompressibility constraint
\begin{equation}
\nabla \cdot \bfu = 0 ; \label{eq:incomp}
\end{equation}
and, of course,
\begin{equation}
\nabla \cdot \bfb = 0 .  \label{eq:divb}
\end{equation}
The external forces $\f_u$
and $\f_b$ are zero-mean, Gaussian random forces that are
uncorrelated with each other and delta correlated in time; their
statistical properties are best described by their covariances in
Fourier space:
\begin{eqnarray}
\langle \hat{f}_{u,m}({\bf k},t)
\hat{f}_{u,n}({\bf k}',t') \rangle &=& 2 D_u k^{-3} {\mathcal
P}_{m,n}({\bf k})\delta({\bf k}+{\bf k}')\delta(t-t'); \nonumber
\\ \langle \hat{f}_{b,m}({\bf k},t) \hat{f}_{b,n}({\bf k}',t')
\rangle &=& 2 D_b k^{-3} {\mathcal P}_{m,n}({\bf k})\delta({\bf
k}+{\bf k}')\delta(t-t'); \label{eq:rfmhdeq}
\end{eqnarray}
here carets denote spatial Fourier transforms, ${\bf k}$ and ${\bf
k}'$ are wave vectors, $k = |{\bf k}|$, the transverse projector
${\mathcal P}_{m,n} \equiv [\delta_{m,n} - (k_m k_n/k^2)]$
ensures that $\nabla \cdot \bfu = 0$ and $\nabla \cdot \bfb = 0$,
and $D_u$ and $D_b$ are measures of the kinetic and magnetic
energy injections~\cite{fournier82}; we do not consider cross
correlations between ${\bf f}_u$ and ${\bf f}_b$ here; and we set
$D_u=D_b$. At the level of our DNS this stochastic forcing is
generated by using the following:
\begin{eqnarray} \hat
f_u(k_x,t) = f_x^u e^{\iota\xi}; ~  \hat f_u(k_y,t) = f_y^u
e^{\iota\xi}; ~ \hat f_u(k_z,t) = f_z^u e^{\iota\xi}; \\ \nonumber
\hat f_b(k_x,t) = f_x^b e^{\iota\xi}; ~  \hat f_b(k_y,t) = f_y^b
e^{\iota\xi}; ~ \hat f_b(k_z,t) = f_z^b e^{\iota\xi}.
\end{eqnarray}
We choose the amplitudes of the isotropic forces as follows:
\begin{eqnarray}
f_x^u= f_{amp}^u \sin\theta \cos\phi;
~ f_y^u= f_{amp}^u \sin\theta \cos\phi; ~ f_z^u= f_{amp}^u
\cos\theta; \\ \nonumber
f_x^u= f_{amp}^b \sin\theta \cos\phi; ~
f_y^b= f_{amp}^b \sin\theta \cos\phi; ~ f_z^b= f_{amp}^b
\cos\theta; 
\end{eqnarray} the amplitudes are
\begin{eqnarray} 
f_{amp}^u &=& f_{ran}^u~k^{-3/2}, \nonumber \\ 
f_{amp}^b &=& f_{ran}^b~k^{-3/2},  \label{eq:famp}
\end{eqnarray} 
where $ f_{ran}^u$ and $ f_{ran}^b$ are Gaussian random numbers.  The angles
$\xi$ and $\theta$ are random numbers distributed uniformly in the interval
$[-~\pi,\pi)$ and the angle $\phi$ is a random number distributed uniformly in
the interval $[0,\pi/2)$. We use a standard pseudospectral
method~\cite{sahoomhd,sahoophd,book-canuto} to solve Eqs.~\ref{eq:incomp}, 
\ref{eq:divb} and \ref{eq:rfmhdeq}, the $2/3$ rule for removing aliasing
errors~\cite{book-canuto}, and an Adams-Bashforth method, with step size
$\delta t$, for time marching~\cite{sahoomhd,sahoophd}.

In our DNSs we use initial (superscript $0$) energy spectra
$E_u^0(k)$ and $E_b^0(k)$, for the velocity and magnetic fields,
respectively, as in Ref.~\cite{sahoomhd,sahoophd}: 
\begin{equation} 
E_u^0(k) = E_b^0(k) = E^0k^4\exp(-2k^2) ;    
\label{einitial}
\end{equation} 
the initial phases of the Fourier components of the velocity and magnetic
fields are different and are chosen such that they are distributed uniformly
between $0$ and $2\pi$.

We have carried out three DNSs,
each with $512^3$ collocation points, with the following three sets of
parameters:
\begin{enumerate}
\item $\nu=10^{-4}$, $\eta=10^{-3}$, ${\rm Pr_M}=0.1$, 
${\rm Re}_\lambda \simeq 200$;
\item $\nu=10^{-3}$, $\eta=10^{-3}$, ${\rm Pr_M}=1$, 
${\rm Re}_\lambda \simeq 50$;
\item $\nu=10^{-3}$, $\eta=10^{-4}$, ${\rm Pr_M}=10$, 
${\rm Re}_\lambda \simeq 40$;
\end{enumerate}
here, ${\rm Re}_\lambda$ denotes the Taylor-microscale Reynolds number (see below)
that we obtain in the statistically steady turbulent state, which is eventually
established because of the stochastic forcing in the 3DRFMHD model.  We then
average the quantities of interest, which are described below, over $15$
velocity- and magnetic-fields configurations that are well-separated in
time.

\subsection{Statistical measures \label{sec:statmeasures}}

To enable a detailed comparison of our results for 3DRFMHD with their
counterparts for 3DMHD, we use the same statistical, measures to characterize
homogeneous, isotropic 3DRFMHD turbulence, that are employed in
Ref.~\cite{sahoomhd,sahoophd}: 
\begin{itemize}
\item The kinetic, magnetic, and total energy spectra $E_u(k) = \sum_{\bfk\ni
|\bfk|=k}|{\tilde\bfu}(\bfk)|^2$, $E_b(k) = \sum_{\bfk\ni
|\bfk|=k}|{\tilde\bfb}(\bfk)|^2$, and $E_T(k) = E_u(k)+E_b(k)$,
respectively. 
\item The kinetic, magnetic, and total energies $E_u = \sum_k
E_u(k)/2$, $E_b = \sum_k E_b(k)/2$, and $E_T = E_u+E_b$. 
\item Spectra for the energy dissipation rates and the effective pressure
$\epsilon_u(k) = \nu k^2 E_u(k)$, $\epsilon_b(k) = \nu k^2 E_b(k)$, and
$P(k) = \sum_{\bfk\ni |\bfk|=k}|\tilde{\bar p}(\bfk)|^2$,
respectively. 
\item The Taylor-microscale Reynolds number ${\rm Re}_{\lambda} =
u_{\rm rms}\lambda/\nu$, the magnetic Taylor-microscale Reynolds number
${\rm Re}_M=u_{\rm rms}\lambda/\eta$, and the magnetic Prandtl
number ${\rm Pr_M}={\rm Re_M}/{\rm Re}=\nu/\eta$, where the
root-mean-square velocity $u_{\rm rms} = \sqrt{2E_u/3}$ and the
Taylor microscale $\lambda = \left[\sum_k k^2E(k)/E\right]^{-1/2}$. 
\item The integral length scale $\ell_I = \left[\sum_k E(k)/k\right]/E$. 
\item The mean kinetic energy dissipation rate per unit mass, $\epsilon_u = \nu
\sum_{i,j} (\partial_iu_j+\partial_ju_i)^2 = \nu\sum_k k^2 E_u(k)$, the
mean magnetic energy dissipation rate per unit mass $\epsilon_b
= \eta \sum_{i,j} (\partial_ib_j+\partial_jb_i)^2 = \eta\sum_k
k^2 E_b(k)$, and the mean energy dissipation rate per unit mass
$\epsilon = \epsilon_u+\epsilon_b$. 
\item The dissipation length scales for velocity and magnetic fields $\eta_d^u
= (\nu^3/\epsilon_u)^{1/4}$ and $\eta_d^b = (\eta^3/\epsilon_b)^{1/4}$,
respectively.  (In all our runs, $k_{\rm max}\eta^u_d \gtrsim
1$ and $k_{\rm max}\eta^b_d \gtrsim 1$, where $k_{\rm max}$ is
the largest wave-vector magnitude allowed by the $2/3$ dealiasing rule.)
\item The eigenvalues $\Lambda_n^u$ and the associated eigenvectors ${\hat
e}_n^u$, with $n=1, 2,$ or $3$, of the rate-of-strain tensor $\mathbb
S$, with components $S_{ij}=\partial_iu_j+\partial_ju_i$;
incompressibility implies $\sum_n\Lambda_n^u = 0$, so at least
one of the $\Lambda_n^u$ must be positive and another negative;
we label them such that $\Lambda_1^u > 0$, $\Lambda_3^u < 0$,
and $\Lambda_2^u$, which can be positive or negative, lies in between them.
\item Probability distribution functions (PDFs): PDFs of $\Lambda_n^u$ and of the
cosines of the angles that the associated eigenvectors make with the
vorticity, the current density $\bfj= \nabla\times\bfb/4\pi$, etc. (These PDFs and 
those of the local cross helicity $H_C=\bfu\cdot\bfb$ quantify the
degree of alignment of pairs of vectors such as $\bfu$ and $\bfb$.)
\item Joint PDFs (JPDFs): We obtain the velocity-derivative tensor $\mathbb A$, with
components $A_{ij}=\partial_iu_j$ and thence the invariants $Q =
-\frac{1}{2}tr(\mathbb{A}^2)$ and $R = -\frac{1}{3}tr(\mathbb{A}^3)$~\cite{cantwell93}, 
which are used to characterize fluid turbulence~\cite{cantwell93,book-davidson,biferale07}. The
zero-discriminant curve $D\equiv \frac{27}{4}R^2+Q^3=0$ and the $Q$ and $R$ axes 
divide the $Q$-$R$ plane into different four regimes: 
\begin{itemize}
\item $Q$ large and negative; vortex formation is not favored because local strains 
are high; also, if $R>0$ ($R<0$), fluid elements experience axial (biaxial) strain.
\item $Q$ large and positive: the flow is predominantly vortical; if, 
in addition, $R>0$ ($R<0$), vortices are stretched (compressed).
\end{itemize}
\item For turbulent fluid flow, contour plots of the JPDF of $Q$ and $R$
(often called $QR$ plots) exhibit a tear-drop shape. We explore these JPDFs and
others, e.g., JPDFs of $\epsilon_u$ and $\epsilon_b$, for 3DRFMHD turbulence.
\item Structure functions: We calculate the order-$p$, longitudinal structure
functions $S_p^a(l) \equiv \langle|\delta a_{\parallel}(\bfx,l)|^p\rangle$; 
the increment of the longitudinal component of the field 
$\bfa$ is $\delta a_{\parallel}(\bfx,l) \equiv
\bfa(\bfx+\bfl,t)-\bfa(\bfx,t)]\cdot\frac{\bfl}{l}$; here, $\bfa$ can be $\bfu$ or 
$\bfb$; we also compute the hyperflatness $F_6^a(l)=S_6^a(l)/[S_2^a(l)]^3$. 
\item Intermittency: For separations $l$ in the inertial range, 
$\eta_d^u, \eta_d^b \ll l \ll L$, we anticipate that
$S_p^a(l)\sim l^{\zeta_p^a}$, with $\zeta_p^a$ the
multiscaling exponents for $\bfa$; the Kolmogorov phenomenology (K41) of
1941~\cite{book-frisch,k41,biskamp00} yields $\zeta_p^{aK41} = p/3$; but 
multiscaling corrections are substantial, with $\zeta_p^a\neq\zeta_p^{aK41}$,
especially for $p \geq 3$ [see Subsec.~\ref{sec:stfn}]. 
\item We also explore the dependence on $l$ of the PDFs of the increments $\delta
a_{\parallel}(\bfx,l)$. 
\end{itemize}

\section{Results \label{sec:results}}

We now present the results of our DNSs for 3DRFMHD with ${\rm Pr_M}=0.1$, ${\rm
Pr_M}=1$, and ${\rm Pr_M}=10$; we also compare these results with their
counterparts for 3DMHD~\cite{sahoomhd,sahoophd}.  Subsection~\ref{sec:tseries}
gives the temporal evolution of the energy, energy-dissipation rates, the
Taylor-microscale Reynolds number, the Taylor microscale, and the integral
length scale; in subsection~\ref{sec:spectra} we present energy,
dissipation-rate, and effective-pressure spectra; subsection~\ref{sec:pdfs} is
devoted to probability distribution functions (PDFs) of quantities such as
components of fields, energy dissipation rates and cosines of angles between
various vectors; in subsection~\ref{sec:stfn} we present  velocity and
magnetic-field structure functions and explain how they can be used to
characterize inertial-range intermittency; subsection~\ref{sec:isosurface}
displays isosurfaces mentioned above; in subsection~\ref{sec:jpdfs} we present
joint PDFs that give statistical correlation between various quantities. 

\subsection{Temporal evolution \label{sec:tseries}}

\begin{figure}[htb]
\begin{center}
\includegraphics[width=0.95\textwidth]{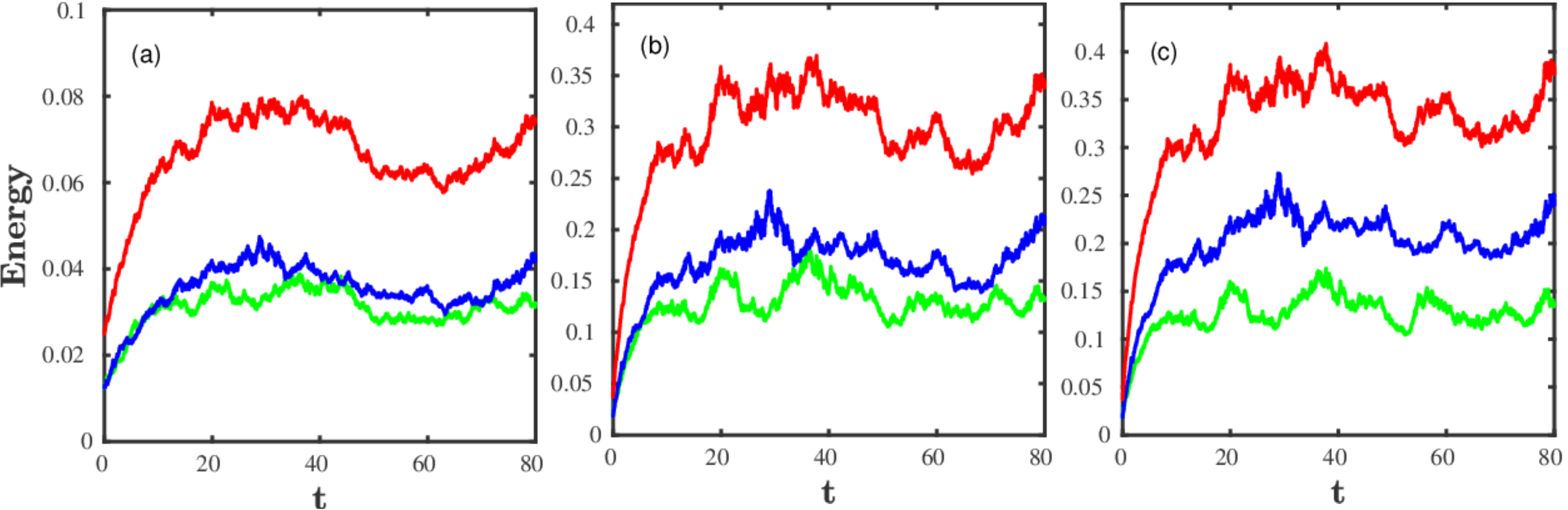}
\end{center}
\caption[]{Plots of the total energy (red line),
kinetic energy (green line), and magnetic energy (blue line) versus time $t$ (product of number of
steps and $\delta t$) for (a) ${\rm Pr_M}=0.1$, (b) ${\rm Pr_M}=1$,
and (c) ${\rm Pr_M}=10$ (cf. Fig. $7$ (b.1), (c.1), and (d.1) in Ref.~\cite{sahoomhd})}.
\label{fig:energy}
\end{figure}

Figures~\ref{fig:energy} (a), (b), and (c) show plots of the total energy $E$
(red line), the total kinetic energy $E_u$ (green line), and the total magnetic
energy $E_b$ (blue line) versus time $t$ for ${\rm Pr_M}=0.1$, ${\rm Pr_M}=1$,
and ${\rm Pr_M}=10$, respectively; these plots are more noisy than their
analogues for statistically steady 3DMHD (cf. Fig. $7$ (b.1), (c.1), and (d.1)
in Ref.~\cite{sahoomhd}). Eventually, a statistically steady state is obtained,
even for 3DRFMHD, in which $E$, $E_u$, and $E_b$ fluctuate about their mean
values $\langle E \rangle$, $\langle E_u \rangle$, and $\langle E_b \rangle$. 

\begin{figure}[htb]
\begin{center}
\includegraphics[width=0.95\textwidth]{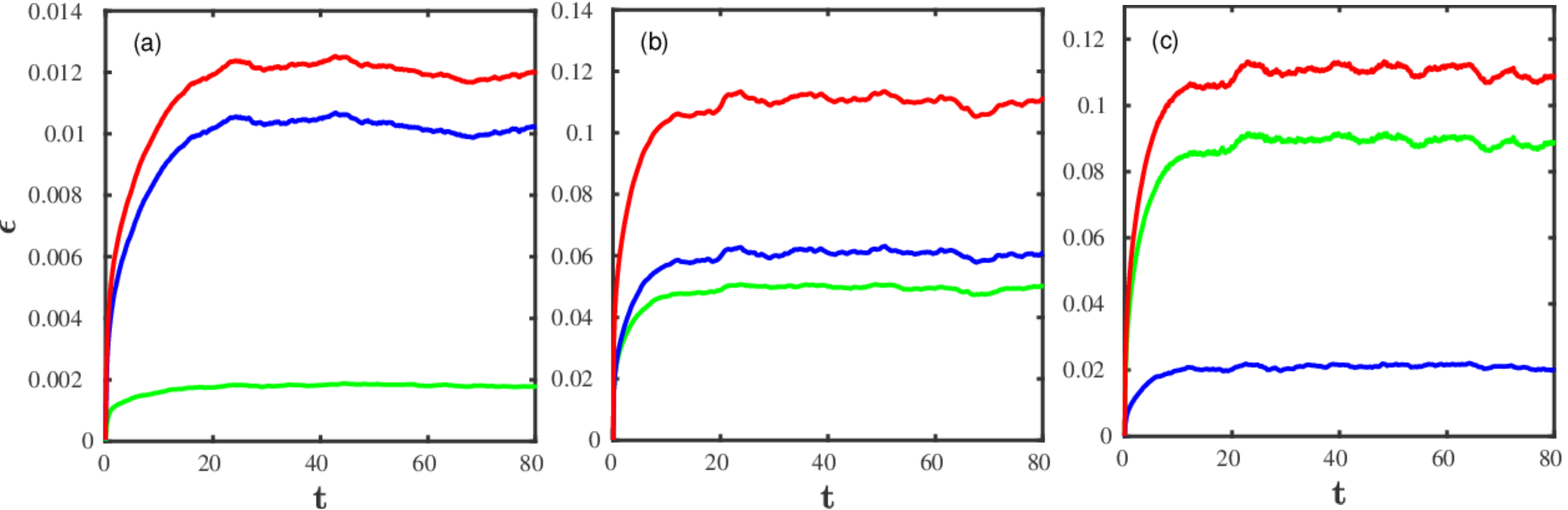}
\end{center}
\caption[]{Plots of the total energy dissipation (red line),
kinetic-energy dissipation (green line) and magnetic-energy
dissipation (blue line) rates versus time $t$ (product of number of
steps and $\delta t$) for (a) ${\rm Pr_M}=0.1$, (b) ${\rm Pr_M}=1$,
and (c) ${\rm Pr_M}=10$ (cf. Fig. $7$ (b.2), (c.2), and (d.2) in Ref.~\cite{sahoomhd})}.
\label{fig:dissipation}
\end{figure}

In Figs.~\ref{fig:dissipation} (a), (b), and (c) we show plots of the total
energy dissipation rate $\epsilon$ (red line), the kinetic-energy
dissipation rate $\epsilon_u$ (green line), and the magnetic-energy
dissipation rate $\epsilon_b$ (blue line) versus $t$ for ${\rm
Pr_M}=0.1$, ${\rm Pr_M}=1$, and ${\rm Pr_M}=10$, respectively. The difference
between the mean values of the kinetic- and magnetic-energy dissipation rates
$(\langle \epsilon_u \rangle- \langle \epsilon_b \rangle)$ is negative at
${\rm Pr_M}=0.1$, it increases when ${\rm Pr_M}=1$, and it finally becomes
positive for ${\rm Pr_M}=0.1$ as we expect; this also occurs in
3DMHD (cf. Fig. $7$ (b.2), (c.2), and (d.2) in Ref.~\cite{sahoomhd}).

\begin{figure}[htb]
\begin{center}
\includegraphics[width=0.95\textwidth]{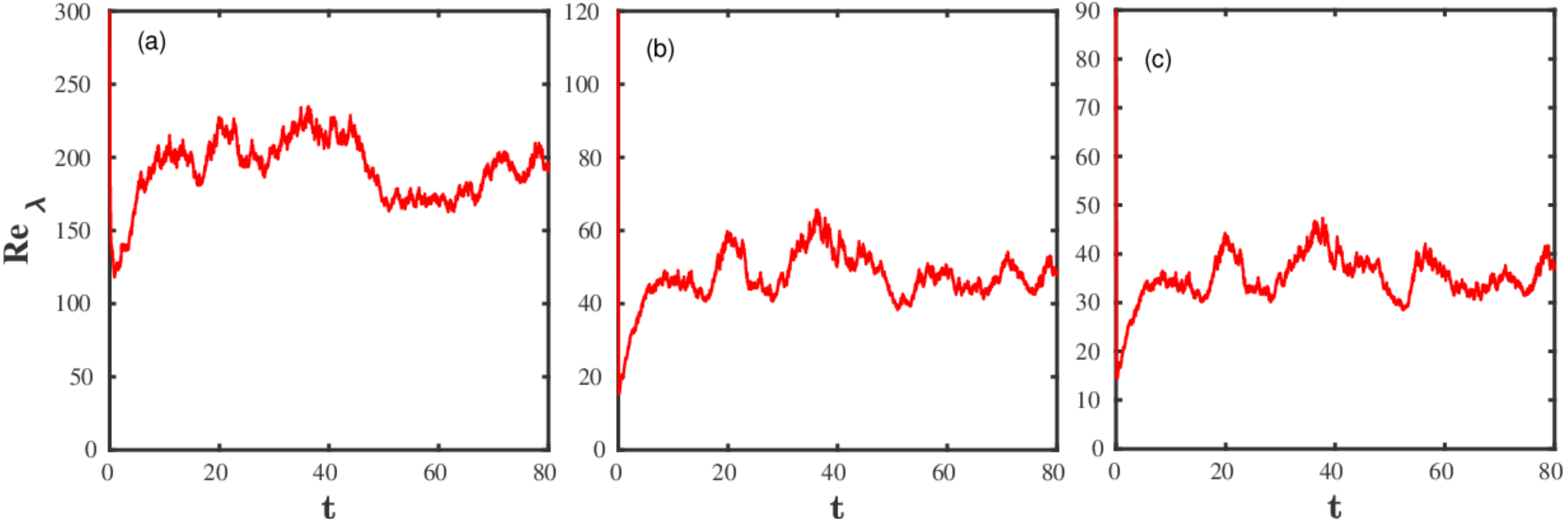}
\end{center}
\caption[]{Plots of the Taylor-microscale Reynolds number ${\rm
Re}_\lambda$ versus time $t$ (product of number of
steps and $\delta t$) for (a) ${\rm Pr_M}=0.1$, (b) ${\rm Pr_M}=1$,
and (c) ${\rm Pr_M}=10$.}
\label{fig:reynolds}
\end{figure}
\begin{figure}[htb]
\begin{center}
\includegraphics[width=0.95\textwidth]{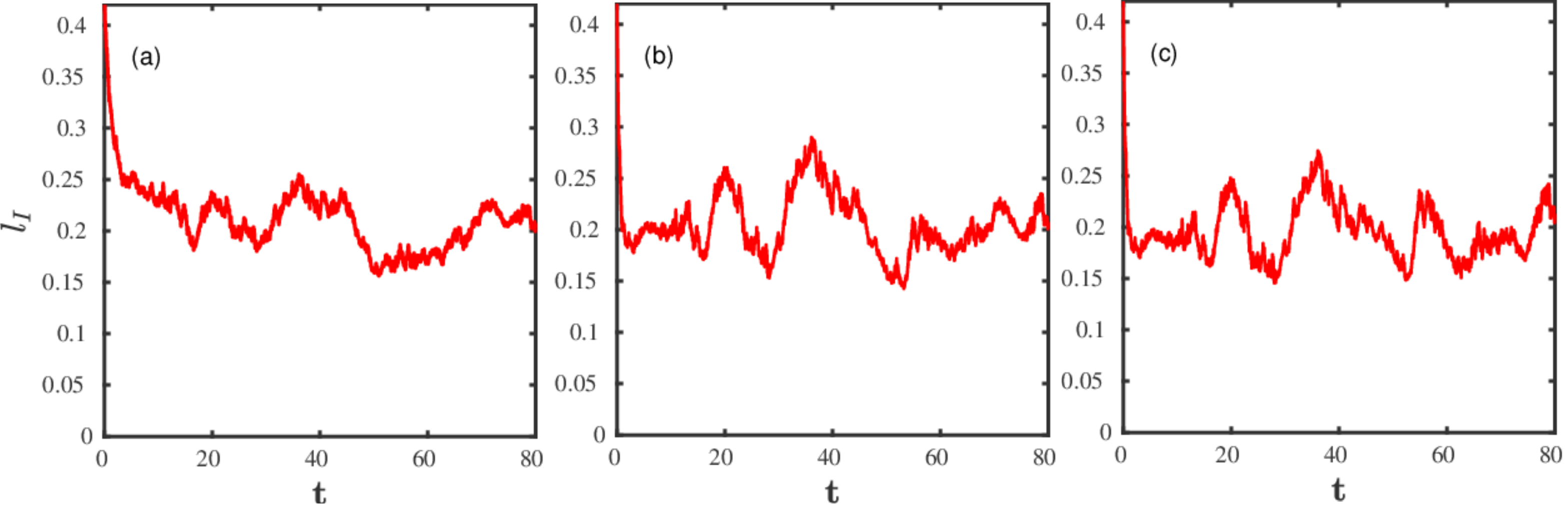}
\end{center}
\caption[]{Plots of the integral scale $\ell_I$ versus time $t$ (product of
number of steps and $\delta t$) for (a) ${\rm Pr_M}=0.1$, (b) ${\rm Pr_M}=1$,
and (c) ${\rm Pr_M}=10$.}
\label{fig:integralscale}
\end{figure}
\begin{figure}[htb]
\begin{center}
\includegraphics[width=0.95\textwidth]{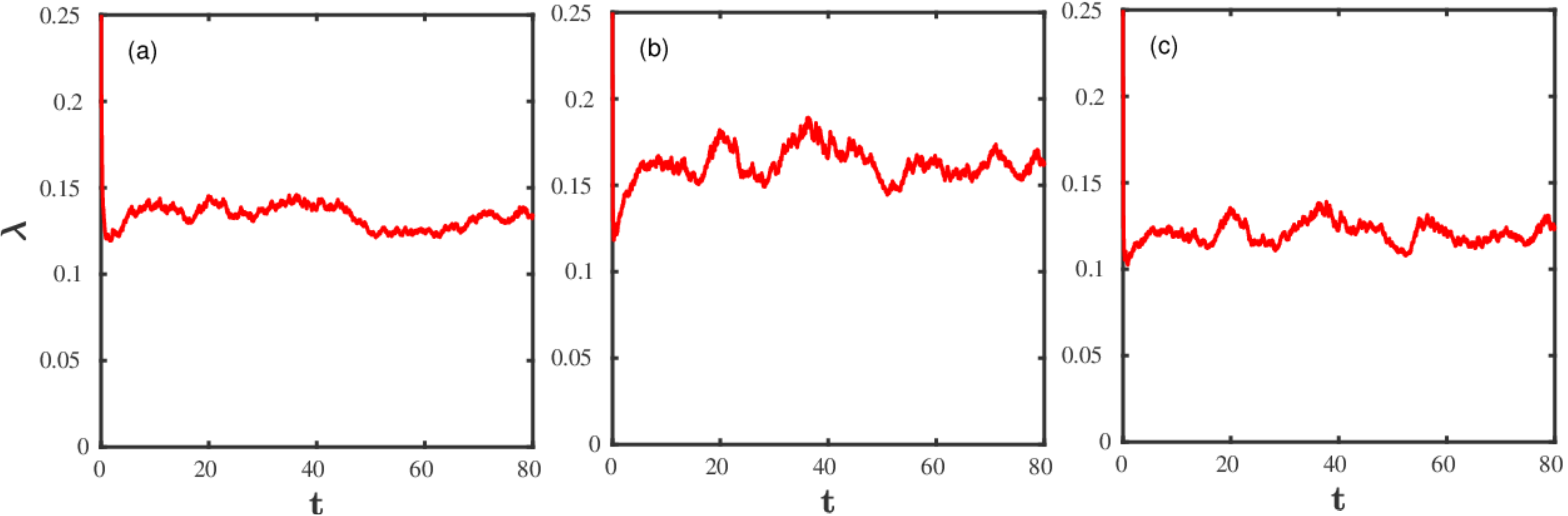}
\end{center}
\caption[]{Plots of the Taylor microscale $\lambda$ versus time $t$ (product
of number of steps and $\delta t$) for (a) ${\rm Pr_M}=0.1$, (b) ${\rm
Pr_M}=1$, and (c) ${\rm Pr_M}=10$.}
\label{fig:tmscale}
\end{figure}

Figures~\ref{fig:reynolds},~\ref{fig:integralscale},~\ref{fig:tmscale} show
the temporal evolution of the Taylor-microscale Reynolds number ${\rm
Re}_\lambda$, the integral scale $\ell_I$, and the Taylor microscale
$\lambda$, respectively, for (a) ${\rm Pr_M}=0.1$, (b) ${\rm Pr_M}=1$, and
(c) ${\rm Pr_M}=10$. Once the statistically steady state is achieved, these quantities
fluctuate about a mean. 
The mean value of ${\rm Re}_\lambda$
decreases with increasing ${\rm Pr_M}$, because we force velocity and
magnetic fields with equal intensity. The mean values of $\ell_I$ and
$\lambda$ do not depend very sensitively on ${\rm Pr_M}$. 

\subsection{Spectra \label{sec:spectra}}
\begin{figure}[htb] \begin{center}
\includegraphics[width=0.95\textwidth]{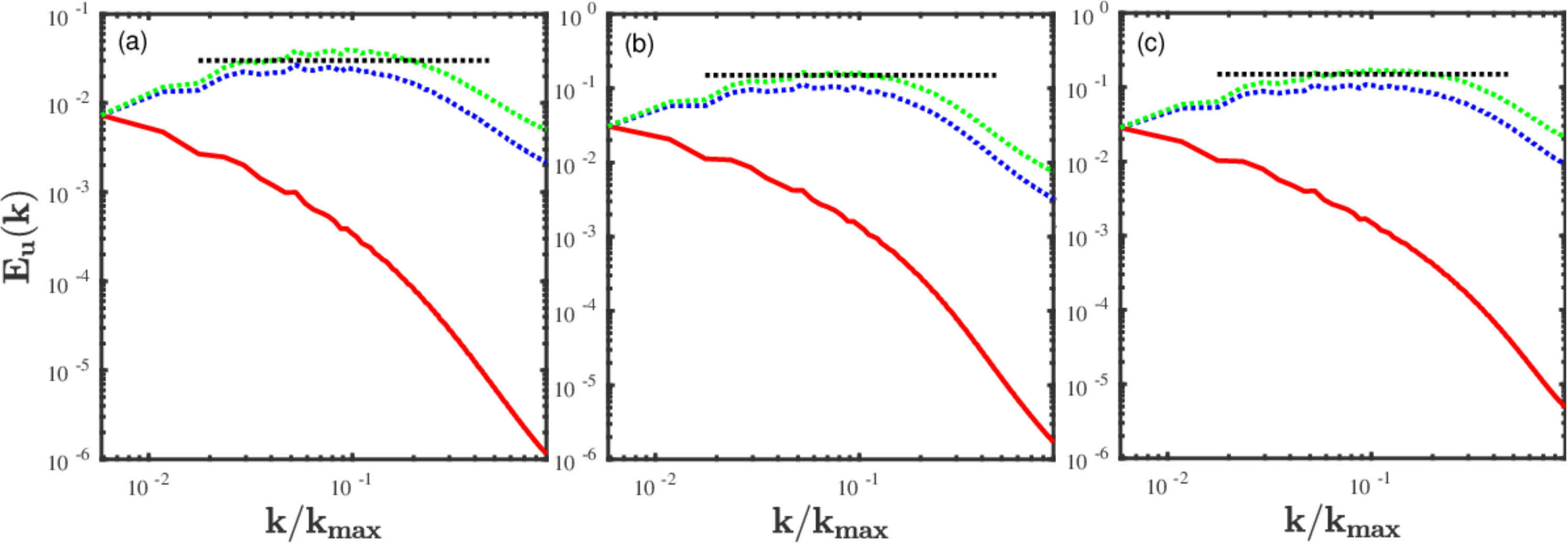}\\
\end{center} 
\caption[]{Log-log (base 10) plots of kinetic energy spectra $E_u(k)$ (red
full line), with $k$, the magnitude of the wave vector and the corresponding
compensated spectra $k^{5/3}E_u(k)$ (green dashed lines) and $k^{3/2}E_u(k)$ (blue dashed lines) 
for (a) ${\rm Pr_M}=0.1$, (b) ${\rm Pr_M}=1$, and (c) ${\rm Pr_M}=10$
(cf. Figs. $8$ (b.3), (c.3), and (d.3) in Ref.~\cite{sahoomhd})}
\label{fig:kinetic-energy-spectra} \end{figure}
\begin{figure}[htb] \begin{center}
\includegraphics[width=0.95\textwidth]{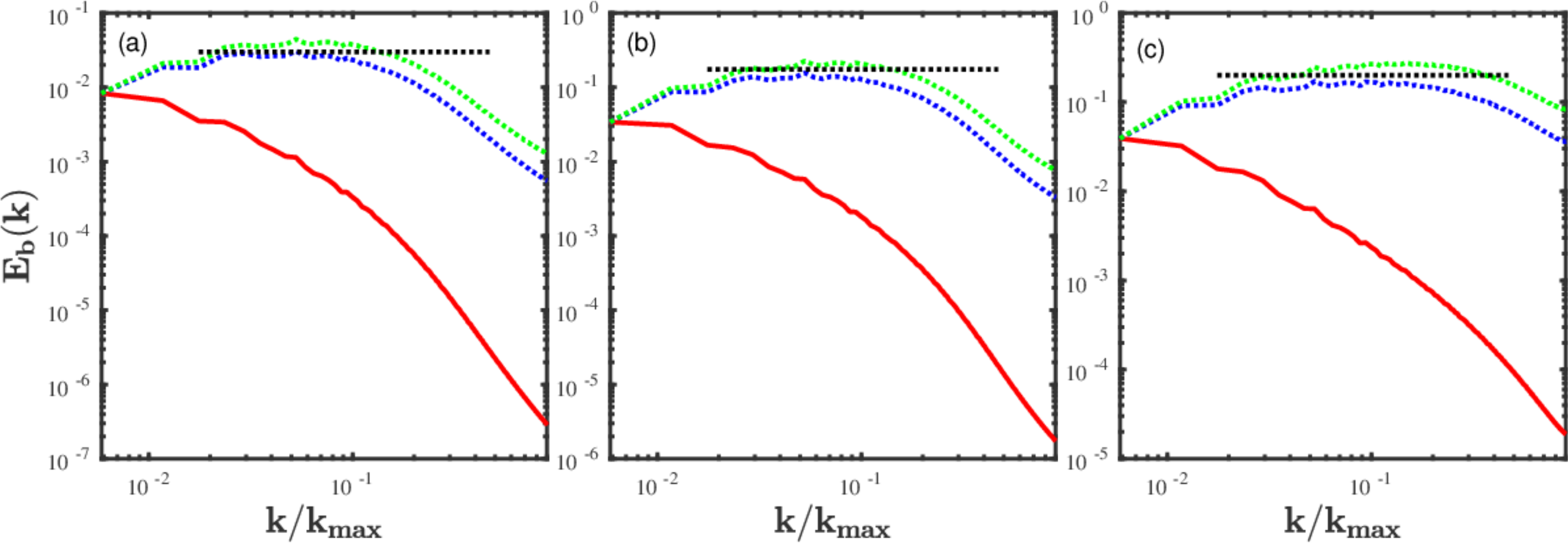}\\
\end{center} 
\caption[]{Log-log (base 10) plots of effective magnetic energy spectra $E_b(k)$ (red
full lines), with $k$, the magnitude of the wave vector and the corresponding
compensated spectra $k^{5/3}E_b(k)$ (green dashed lines) and $k^{3/2}E_b(k)$ (blue dashed lines) for (a) ${\rm Pr_M}=0.1$, (b) ${\rm Pr_M}=1$, and (c) ${\rm Pr_M}=10$
(cf. Figs. $8$ (b.3), (c.3), and (d.3) in Ref.~\cite{sahoomhd}).}
\label{fig:magnetic-energy-spectra} \end{figure}

Figures~\ref{fig:kinetic-energy-spectra} (a), (b), and (c) show plots of the
kinetic-energy spectra $E_u(k)$ (red full line), and the corresponding
compensated spectra $k^{5/3}E_u(k)$ (green dashed line) and $k^{3/2}E_u(k)$
(blue dashed line) for ${\rm Pr_M}=0.1$, ${\rm Pr_M}=1$, and ${\rm Pr_M}=10$,
respectively; figures~\ref{fig:magnetic-energy-spectra} (a), (b), and (c) show
plots of the magnetic energy spectra $E_b(k)$ (red full line), and the
corresponding compensated spectra $k^{5/3}E_b(k)$ (green dashed line) and
$k^{3/2}E_b(k)$ (blue dashed line) for ${\rm Pr_M}=0.1$, ${\rm Pr_M}=1$, and
${\rm Pr_M}=10$, respectively, for 3DRFMHD. 
For ${\rm Pr_M}=1$, both kinetic- and magnetic-energy spectra have comparable
inertial ranges; the former shows a larger (smaller) dissipation range than the
latter as the magnetic Prandtl number increases (decreases) to ${\rm Pr_M}=10$
(to ${\rm Pr_M}=0.1$). These results are consistent the inertial-range
behaviors of energy spectra in 3DMHD (cf. Figs. $8$ (b.3), (c.3), and (d.3) in
Ref.~\cite{sahoomhd}). 

Note that, because of our power-law forcing, these energy spectra for 3DRFMHD
fall more slowly in the dissipation range than their counterparts in 3DMHD; the
latter decay as $k^{\gamma}\exp(-\delta k)$ in the deep dissipation
range~\cite{sahoomhd,sahoophd}; such an exponential decay can be obtained for
3DRFMHD, and other randomly forced models with such
forcing~\cite{mitra05,sain98,biferale04}, only if we introduce an ultraviolet
cutoff for $k$ in the forcing term. Furthermore, for 3DRFMHD we expect mild
logarithmic corrections to the inertial-range, power-law forms of these spectra
in 3DMHD as argued, e.g., for the analogous 3D Navier-Stokes~\cite{sain98} and
1D Burgers~\cite{mitra05} equations with power-law, random forcing; such subtle
logarithmic corrections cannot be uncovered reliably given the spatial
resolution of our 3DRFMHD runs. These comments also apply to the inertial- and
dissipation-range forms of the other spectra we discuss below.

\begin{figure}[htb] \begin{center}
\includegraphics[width=0.95\textwidth]{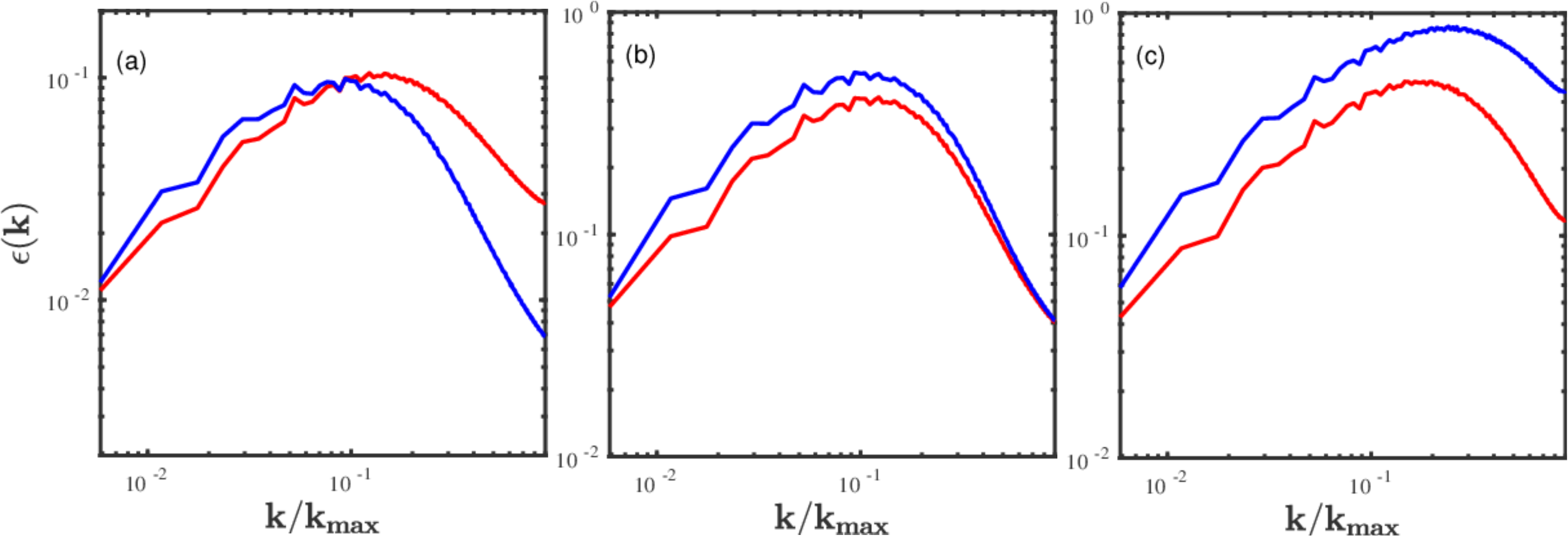}
\end{center} 
\caption[]{Log-log (base 10) plots of energy-dissipation spectra for the
velocity (red lines) and magnetic (blue lines) fields, with $k$
the magnitude of the wave vector for (a) ${\rm Pr_M}=0.1$, (b) ${\rm
Pr_M}=1$, and (c) ${\rm Pr_M}=10$  (cf. Figs. $10$ (b.3),
(c.3), and (d.3) in Ref.~\cite{sahoomhd}).} 
\label{fig:dissipation-spectra}
\end{figure}

In figures~\ref{fig:dissipation-spectra} (a), (b), and (c) we present the
energy-dissipation spectra for the velocity (red lines) and magnetic (blue
lines) fields for ${\rm Pr_M}=0.1$, ${\rm Pr_M}=1$, and ${\rm Pr_M}=10$,
respectively.  These spectra have maxima, as in 3DMHD (cf. Figs. $10$ (b.3),
(c.3), and (d.3) in Ref.~\cite{sahoomhd}), at the value of $k$ at which the
dissipation becomes significant and the dissipation range sets in; given that
we have such maxima in all our dissipation spectra, we conclude that we have
resolved both fluid and magnetic dissipation ranges adequately for all the
values of ${\rm Pr_M}$ that we use. 

\begin{figure}[htb] \begin{center}
\includegraphics[width=0.95\textwidth]{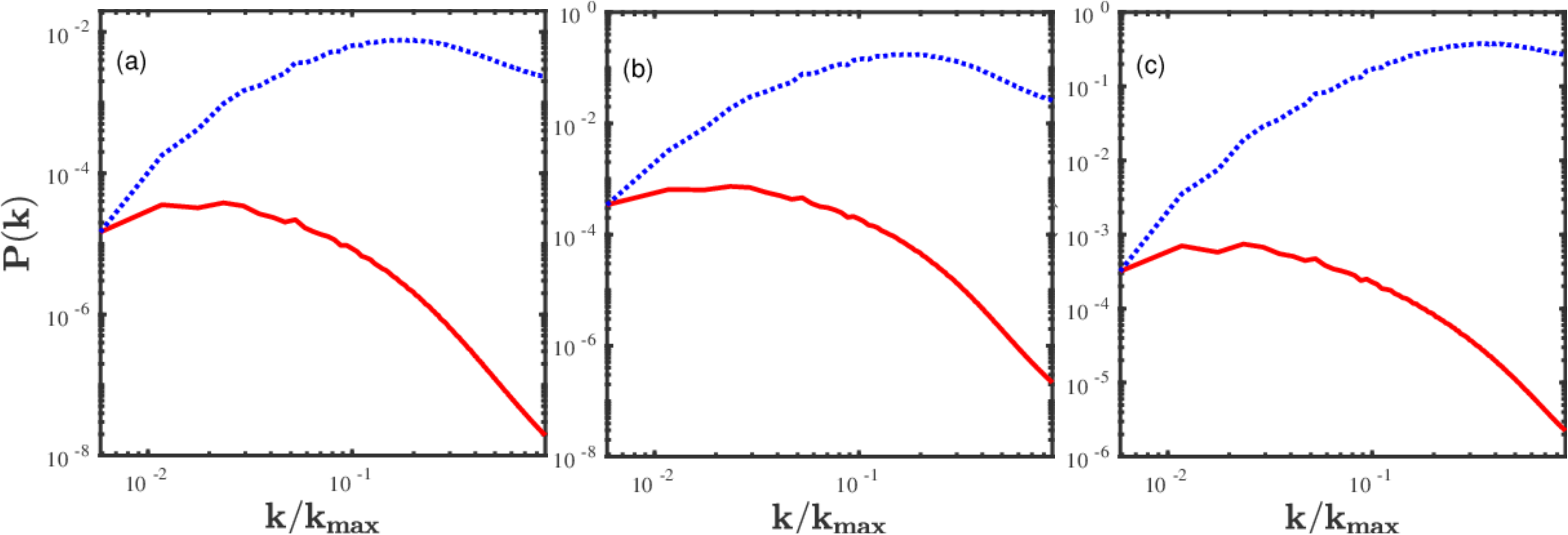}\\
\end{center} 
\caption[]{Log-log (base 10) plots of effective pressure spectra $P(k)$ (red
full lines), with $k$ the magnitude of the wave vector and the corresponding
compensated spectra $k^{7/3}P(k)$ (blue dashed lines) for (a) ${\rm
Pr_M}=0.1$, (b) ${\rm Pr_M}=1$, and (c) ${\rm Pr_M}=10$ (cf. Figs. $12$ (b.2), (c.2), and (d.2) in Ref.~\cite{sahoomhd})}.
\label{fig:pressure-spectra} \end{figure}

Spectra for the effective pressure $P(k)$ (red full lines) and their
compensated versions $k^{7/3}P(k)$ (blue dashed lines) are shown in
figures~\ref{fig:pressure-spectra} (a), (b), and (c) for ${\rm Pr_M}=0.1$, ${\rm
Pr_M}=1$, and ${\rm Pr_M}=10$, respectively. The compensated spectra show that,
for all our runs, the inertial-range behaviors of these effective-pressure
spectra are consistent with the K41 power law $k^{-7/3}$ that goes with the
$k^{-5/3}$ behaviors of the energy spectra discussed above (cf. Figs. $12$
(b.2), (c.2), and (d.2), for 3DMHD, in Ref.~\cite{sahoomhd}).

\subsection{Probability distribution functions \label{sec:pdfs}}

In this subsection we use the results of our DNS studies, for ${\rm Pr_M}=0.1$,
${\rm Pr_M}=1$, and ${\rm Pr_M}=10$, to obtain several probability distribution
functions (PDFs) that characterize the statistical properties of 3DRFMHD
turbulence; we also compare them with their 3DMHD counterparts in Ref.~\cite{sahoomhd}.

\begin{figure}[htb]
\begin{center}
\includegraphics[width=0.95\textwidth]{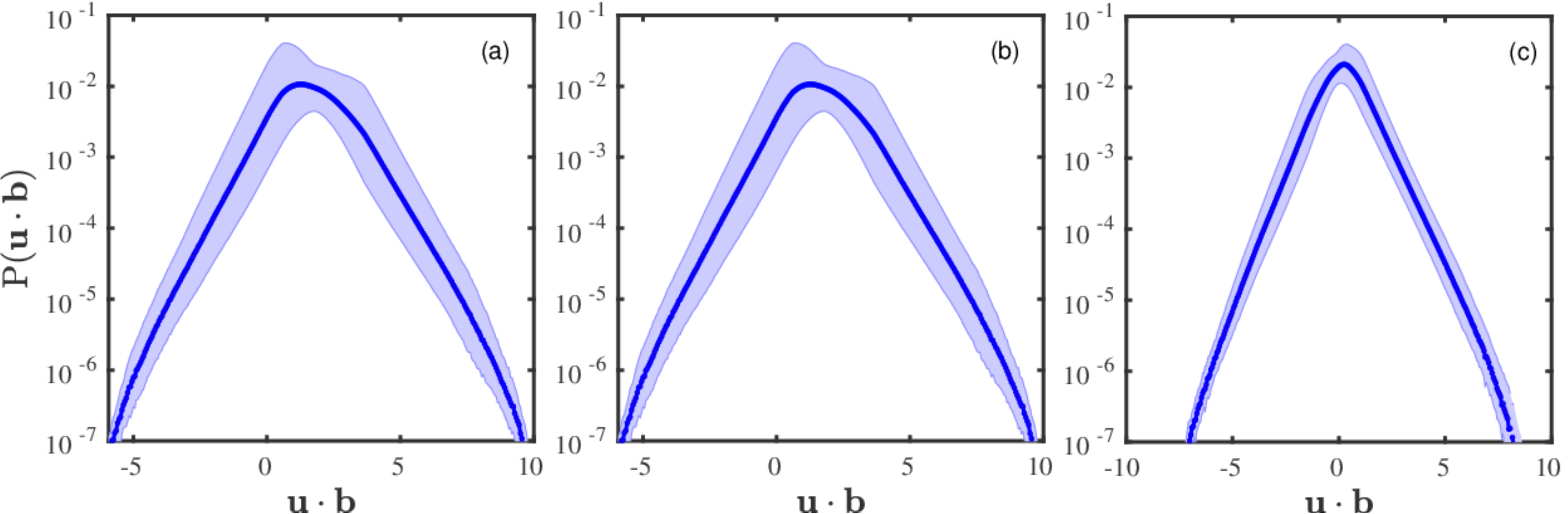}
\end{center}
\caption{Semilog (base 10) plots of PDFs of the moduli of the cross helicity
$H_C = \bfu\cdot\bfb$ for (a) ${\rm Pr_M}=0.1$, (b) ${\rm Pr_M}=1$, and (c)
${\rm Pr_M}=10$, with the arguments of the PDFs scaled by their root-mean-square values. 
One-standard-deviation error bars are indicated by the shaded regions 
(cf. Figs. $18$ (b.3), (c.3), and (d.3), for 3DMHD, in Ref.~\cite{sahoomhd}).}
\label{fig:vdotb}
\end{figure}
\begin{figure}[htb]
\begin{center}
\includegraphics[width=0.95\textwidth]{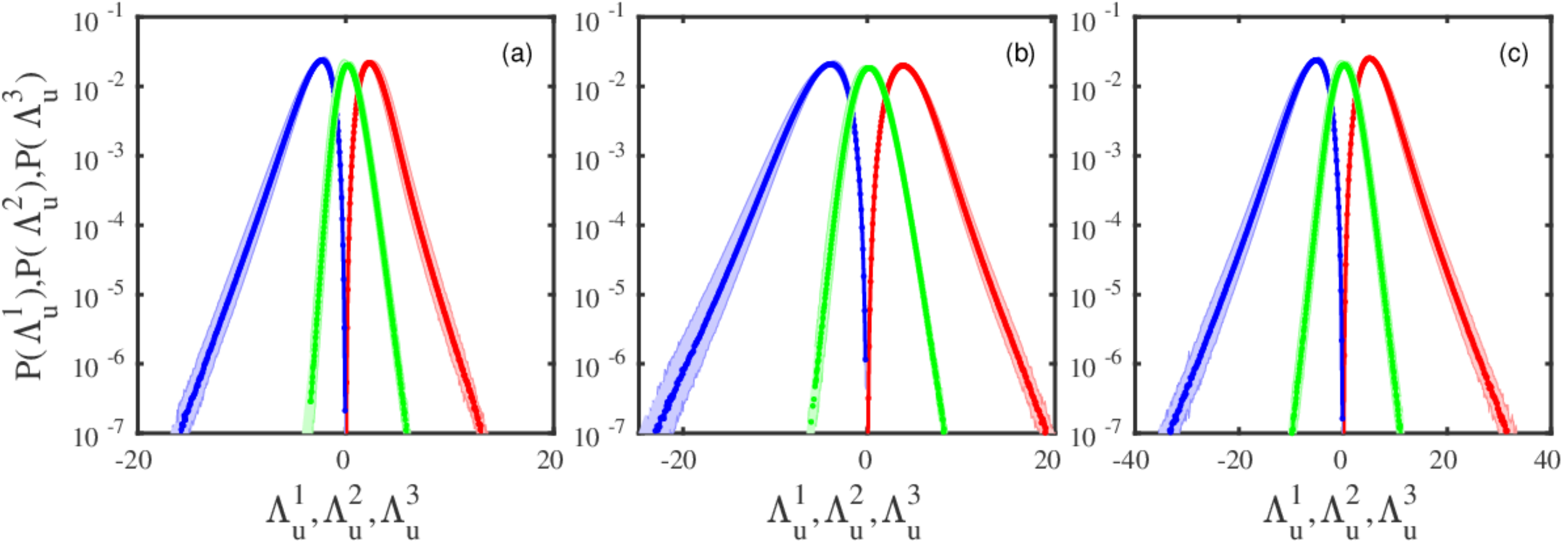}
\end{center}

\caption[] {Semilog (base 10) plots of PDFs of eigenvalues $\Lambda^1_u$ (red
lines), $\Lambda^2_u$ (green lines), and $\Lambda^3_u$ (blue lines) of
the rate-of-strain tensor $\mathbb S$; for (a) ${\rm Pr_M}=0.1$, (b)
${\rm Pr_M}=1$, and (c) ${\rm Pr_M}=10$.  Arguments of PDFs are scaled
by their root-mean-square values. One-standard-deviation error bars are
indicated by the shaded regions (Fig. $19$ in Ref.~\cite{sahoomhd}
gives these PDFs for 3DMHD, but only for the case of decaying 3D MHD turbulence).}

\label{fig:lambda-u}
\end{figure}
\begin{figure}[htb]
\begin{center}
\includegraphics[width=\textwidth]{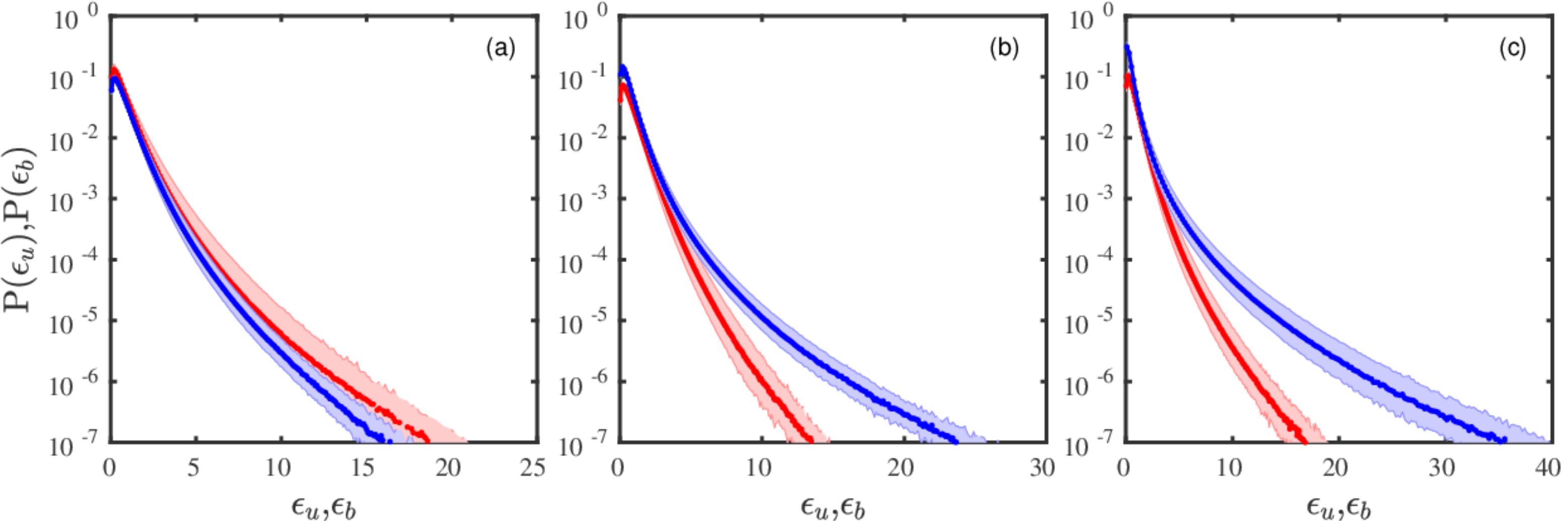}
\end{center}
\caption[]{Semilog (base 10) plots of PDFs of the local kinetic-energy
dissipation rate $\epsilon_u$ (red lines) and the magnetic-energy
dissipation rate $\epsilon_b$ (blue lines), with the arguments scaled by
their root-mean-square values for (a) ${\rm Pr_M}=0.1$, (b) ${\rm Pr_M}=1$, and
(c) ${\rm Pr_M}=10$. One-standard-deviation error bars are indicated by the shaded 
regions (cf. Figs. $20$ (b.3), (c.3), and (d.3), for 3DMHD, in Ref.~\cite{sahoomhd}).}
\label{fig:epsilon}
\end{figure}
\begin{figure}[htb]
\begin{center}
\includegraphics[width=0.95\textwidth]{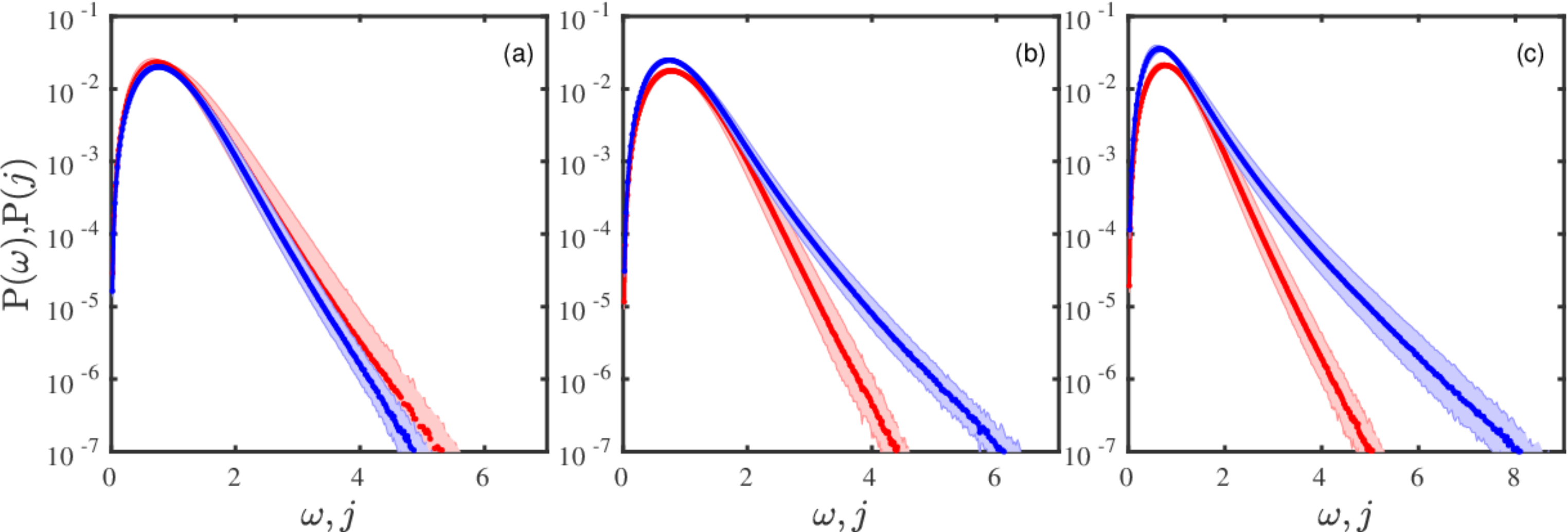}
\end{center}

\caption{Semilog (base 10) plots of PDFs of the moduli of the local vorticity
(red lines) and the current density (blue lines), $\omega$ and $j$,
respectively, with the arguments of the PDFs scaled by their
root-mean-square values for (a) ${\rm Pr_M}=0.1$, (b) ${\rm Pr_M}=1$,
and (c) ${\rm Pr_M}=10$. One-standard-deviation error bars are
indicated by the shaded regions (cf. Figs. $21$ (b.3), (c.3), and (d.3), for 3DMHD, in
Ref.~\cite{sahoomhd}).}

\label{fig:wj}
\end{figure}
\begin{figure}[htb]
\begin{center}
\includegraphics[width=\textwidth]{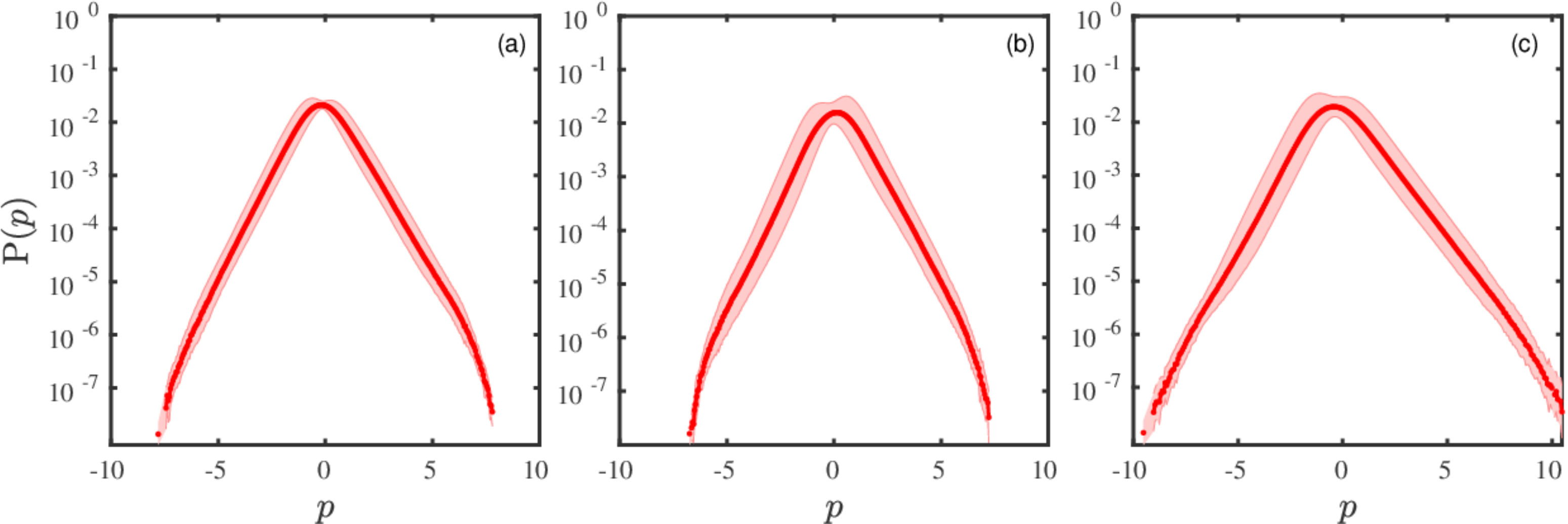}\\
\end{center}

\caption[]{Semilog (base 10) plots of PDFs of local effective pressure
fluctuations (red lines), with the arguments of the PDFs scaled by
their root-mean-square values, for (a) ${\rm Pr_M}=0.1$, (b) ${\rm
Pr_M}=1$, and (c) ${\rm Pr_M}=10$. One-standard-deviation error bars
are indicated by the shaded regions (cf. Figs. $22$ (b.2), (c.2), and (d.2), 
for 3DMHD, in Ref.~\cite{sahoomhd}).}

\label{fig:pressure}
\end{figure}

Probability distribution functions of the modulus of the local cross helicity
$H_C = \bfu\cdot\bfb$ are shown in figures~\ref{fig:vdotb} (a), (b), and (c) for
${\rm Pr_M}=0.1$, ${\rm Pr_M}=1$, and ${\rm Pr_M}=10$, respectively; the
arguments of these PDFs are scaled by their root-mean-square values. For all
the values of ${\rm Pr_M}$ that we consider, these PDFs are (a) peaked very
close to the mean value $\mu_{H_C} \simeq 0$, (b) distinctly non-Gaussian, and
(c) negatively skewed. The mean $\mu_{H_C}$, standard deviation $\sigma_{H_C}$,
skewness $\gamma_{3, H_C}$, and kurtosis $\gamma_{4, H_C}$ of the cross
helicity are given in Table~\ref{table:hc}. The result $\mu_{H_C} \simeq 0$
indicates a strong tendency for the alignment or antialignment of $\bfu$ and
$\bfb$ in 3DRFMHD, a property it shares with 3DMHD (cf. bottom row of Figs.
$18$ (b.3), (c.3), and (d.3), for 3DMHD, in Ref.~\cite{sahoomhd}); of course, this alignment or
antialignment is not perfect as can be seen from the broad PDF of $H_C$.

The PDFs of the eigenvalues $\Lambda^1_u$ (red lines), $\Lambda^2_u$ (green
lines), and $\Lambda^3_u$ (blue lines) of the rate-of-strain tensor $\mathbb S$
are shown in figures~\ref{fig:lambda-u} (a), (b), and (c) for ${\rm Pr_M}=0.1$,
${\rm Pr_M}=1$, and ${\rm Pr_M}=10$. [We label the eigenvalues such that
$\Lambda^1_u > \Lambda^2_u > \Lambda^3_u$.] These PDFs for 3DRFMHD are
qualitatively similar to their fluid-turbulence or
3DMHD~\cite{sahoomhd,sahoophd} analogues. (Figure $19$ in Ref.~\cite{sahoomhd}
gives these PDFs for 3DMHD, but only for the case of decaying 3D MHD
turbulence, so a direct comparison is not possible.) We recall that these
eigenvalues provide measures of the local stretching and compression of the
fluid. 

In figures~\ref{fig:epsilon} (a), (b), and (c) we show PDFs of the
kinetic-energy dissipation rate $\epsilon_u$ (red lines) and the
magnetic-energy dissipation rate $\epsilon_b$ (blue lines) for ${\rm
Pr_M}=0.1$, ${\rm Pr_M}=1$, and ${\rm Pr_M}=10$.  All these PDFs have long
tails; the tail of the PDF for $\epsilon_b$ extends further than its
counterpart for $\epsilon_u$ as the magnetic Prandtl number is increased;
these PDFs are similar to their counterparts for 3DMHD~\cite{sahoomhd,sahoophd}
(cf. Figs. $20$ (b.3), (c.3), and (d.3), for 3DMHD, in Ref.~\cite{sahoomhd}). The
tails of these PDFs indicates that large values of $\epsilon_b$ are more
likely to appear than large values of $\epsilon_u$ and, given the long tails
of these PDFs, suggests that, except at the smallest values of ${\rm Pr_M}$
we have used, we might obtain more marked intermittency for the magnetic
field than for the velocity field. The means $\mu_{\epsilon_u}$ and
$\mu_{\epsilon_b}$, standard deviations $\sigma_{\epsilon_u}$ and
$\sigma_{\epsilon_b}$, skewnesses $\gamma_{3, \epsilon_u}$ and $\gamma_{3,
\epsilon_b}$, and kurtoses  $\gamma_{4, \epsilon_u}$ and $\gamma_{4,
\epsilon_b}$ of the local dissipation rates are given in
Table~\ref{table:epsilon}. 

Similar trends emerge if we examine the PDFs of the moduli of the vorticity
and the current density $\omega$ (red lines) and $j$ (blue lines), respectively. 
These are shown in figures~\ref{fig:wj} (a), (b), and (c)
for ${\rm Pr_M}=0.1$, ${\rm Pr_M}=1$, and (c) ${\rm Pr_M}=10$. The tail of
the PDF for $j$ extends further than the tail of that for $\omega$ as we
increase the Prandtl number, which again indicates more intermittency for the
magnetic field than for the velocity field with increasing ${\rm Pr_M}$.
These PDFs are also similar to their counterparts for 3DMHD~\cite{sahoomhd,sahoophd}
(cf. Figs. $21$ (b.3), (c.3), and (d.3), for 3DMHD, in Ref.~\cite{sahoomhd}).
The means $\mu_{\omega}$ and $\mu_{j}$, standard deviations $\sigma_{\omega}$
and $\sigma_{j}$, skewnesses $\gamma_{3, \omega}$ and $\gamma_{3, j}$, and
kurtoses $\gamma_{4, \omega}$ and $\gamma_{4, j}$ of the moduli of local
vorticity and current density are given in Table~\ref{table:omega}. 

Probability distribution functions of the local effective pressure (red lines)
are shown in figures~\ref{fig:pressure} (a), (b), and (c)  for ${\rm Pr_M}=0.1$,
${\rm Pr_M}=1$, and ${\rm Pr_M}=10$, respectively. As in 3DMHD (cf. Figs. $22$
(b.2), (c.2), and (d.2) in Ref.~\cite{sahoomhd}), these PDFs for 3DRFMHD are
(a) distinctly non-Gaussian, (b) peaked at the $\mu_p = 0$, and (c) are
positively skewed, for ${\rm Pr_M}=1$, ${\rm Pr_M}=10$, and ${\rm Pr_M}=0.1$.
The mean $\mu_{p}$, standard deviation $\sigma_{p}$, skewness $\gamma_{3, p}$,
and kurtosis $\gamma_{4, p}$ of the effective pressure are given in
Table~\ref{table:pressure}.

\begin{table}
\caption{The mean $\mu_{H_c}$, standard deviation $\sigma_{H_c}$,  skewness
$\mu_{3,H_C}$, and kurtosis $\mu_{4,H_C}$ of the PDF of the cross helicity
${H_C}$ for our runs for 3DRFMHD; columns 6 and 7 give, respectively, the
mean energy $E$ and ratio of the means of the cross helicity and the energy,
i.e., $\mu_{H_c}/E$.}
\label{table:hc}
\begin{center}
\begin{tabular}{lllllll}
\hline\noalign{\smallskip}
${\rm Pr_M}$ & $\mu_{H_C}$~~~ & $\sigma_{H_C}$~~~ & $\gamma_{3,H_C}$ &
$\gamma_{4,H_C}$ & $E$ & $\mu_{H_C}/E$ \\
\noalign{\smallskip}\hline\noalign{\smallskip}
0.1 & 0.04 & 0.03 & 0.30 & 3.00 & 0.07 & 0.58 \\
1   & 0.16 & 0.14 & 0.30 & 3.00 & 0.31 & 0.52 \\
10  & 0.02 & 0.12 & 0.02 & 4.13 & 0.35 & 0.06 \\
\noalign{\smallskip}\hline
\end{tabular}
\end{center}
\end{table}

\begin{table}
\caption{The mean $\mu$, standard deviation $\sigma$, 
skewness $\gamma_3$, and kurtosis $\gamma_4$ of
the PDFs of the modulus of the local vorticity $\omega$ and local current
density $j$.}
\label{table:omega}
\begin{center}
\begin{tabular}{lllll|llll}
\hline\noalign{\smallskip}
${\rm Pr_M}$ & $\mu_{\omega}$~~~ & $\sigma_{\omega}$~~~ & $\gamma_{3,\omega}$~~~ & $\gamma_{4,\omega}$
& $\mu_{j}$~~~ & $\sigma_{j}$~~~ & $\gamma_{3,j}$~~~ & $\gamma_{4,j}$\\
\noalign{\smallskip}\hline\noalign{\smallskip}
0.1 & 2.66 & 1.30 & 1.01 & 4.90 & 2.08 & 0.96 & 0.89 & 4.49 \\
1   & 4.61 & 2.09 & 0.78 & 3.99 & 5.06 & 2.45 & 1.16 & 5.97 \\
10  & 5.52 & 2.59 & 0.92 & 4.59 & 8.28 & 4.69 & 1.79 & 9.64 \\
\noalign{\smallskip}\hline
\end{tabular}
\end{center}
\end{table}

\begin{table}
\caption{The mean $\mu$, standard deviation $\sigma$, 
skewness $\gamma_3$, and kurtosis $\gamma_4$ of
the PDFs of the modulus of the local energy dissipation rates
$\epsilon_u$ and $\epsilon_b$.}
\label{table:epsilon}
\begin{center}
\begin{tabular}{lllll|llll}
\hline\noalign{\smallskip}
${\rm Pr_M}$ & $\mu_{\epsilon_u}$~~~ & $\sigma_{\epsilon_u}$~~~ &
$\gamma_{3,\epsilon_u}$~~~ & $\gamma_{4,\epsilon_u}$
& $\mu_{\epsilon_b}$~~~ & $\sigma_{\epsilon_b}$~~~ &
$\gamma_{3,\epsilon_b}$~~~ & $\gamma_{4,\epsilon_b}$\\
\noalign{\smallskip}\hline\noalign{\smallskip}
0.1 & 1.7 $\times 10^{-3}$ &  1.7$\times 10^{-3}$ & 3.21 & 24.85 & 1.0$\times 10^{-2}$ & 1.0$\times 10^{-2}$ & 2.86 & 20.16 \\
1   & 5.0 $\times 10^{-2}$ &  4.7$\times 10^{-2}$ & 2.45 & 14.48 & 6.1$\times 10^{-2}$ & 6.6$\times 10^{-2}$ & 4.17 & 45.06 \\
10  & 7.2 $\times 10^{-2}$ &  7.1$\times 10^{-2}$ & 2.96 & 21.50 & 1.7$\times 10^{-2}$ & 2.5$\times 10^{-2}$ & 6.84 & 110.8 \\
\noalign{\smallskip}\hline
\end{tabular}
\end{center}
\end{table}

\begin{table}
\caption{The mean $\mu_{p}$, standard deviation 
$\sigma_{p}$, skewness $\gamma_{3,p}$, and kurtosis 
$\gamma_{4,p}$ of the PDFs of the local effective pressure $\bar{p}$.}
\label{table:pressure}
\begin{center}
\begin{tabular}{lllll}
\hline\noalign{\smallskip}
${\rm Pr_M}$ & $\mu_{p}$~~~ & $\sigma_{p}$~~~ & 
$\gamma_{3,p}$~~~ & $\gamma_{4,p}$\\
\noalign{\smallskip}\hline\noalign{\smallskip}
0.1 & -7$\times 10^{-3}$ & 0.05 & 0.19 & 4.12 \\
1   &  4$\times 10^{-3}$ & 0.05 & 0.03 & 3.49 \\
10  & -2$\times 10^{-2}$ & 0.08 & 0.34 & 3.85 \\
\noalign{\smallskip}\hline
\end{tabular}
\end{center}
\end{table}

\subsubsection{PDFs of velocity- and magnetic-field gradients}

To examine closely the nature of intermittency in 3DRFMHD we study the
length-scale dependence of PDFs of increments of velocity and magnetic fields
of the form $\delta a_{\parallel}(\bfx,l) \equiv
\bfa(\bfx+\bfl,t)-\bfa(\bfx,t)]\cdot\frac{\bfl}{l}$, with $\bfa$ either
$\bfu$ or $\bfb$, $l = |\bfl|$ the length scale, and $\bfx$ an origin over
which we can average to determine the dependence of the PDFs of $\delta
a_{\parallel}$ on the scale $l$; for notational convenience, such velocity
and magnetic-field increments are denoted by $\delta u(l)$ and $\delta b(l)$
in our plots.

\begin{figure}[htb]
\begin{center}
\includegraphics[width=0.95\textwidth]{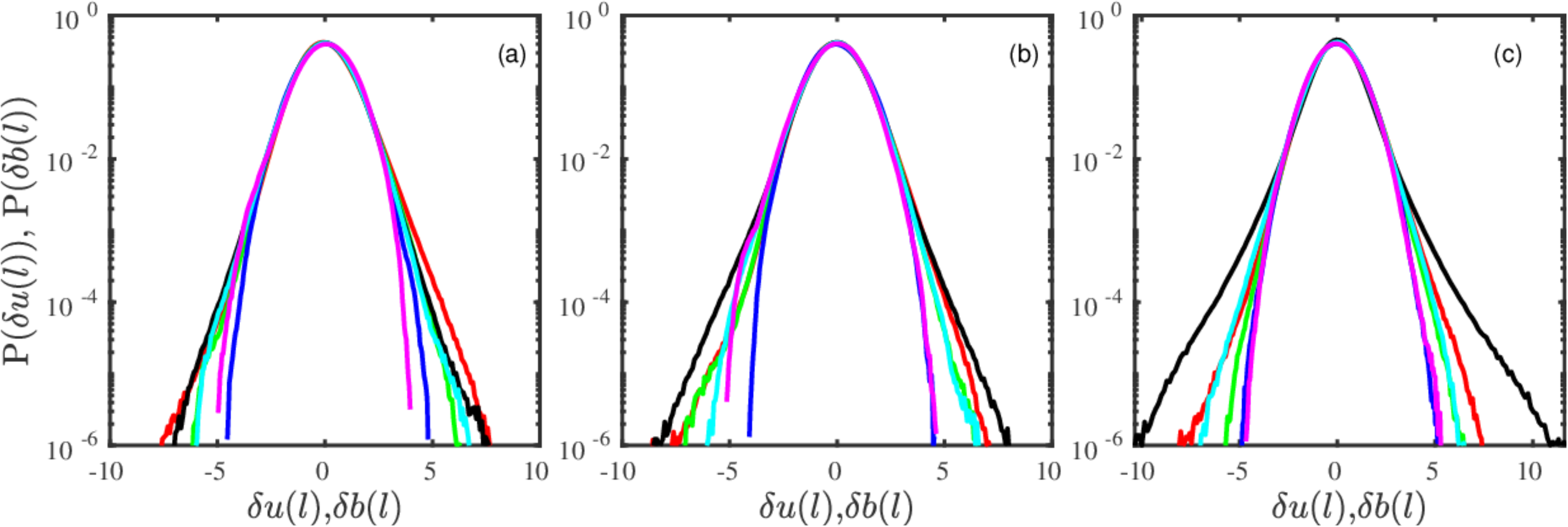}\\
\end{center}
\caption[]{Semilog (base 10) plots of PDFs of velocity increments $\delta
u(l)$, for separations $l = 2\delta x$ (red lines), $10\delta x$ (green
lines), and $100\delta x$ (blue lines), and of magnetic-field
increments $\delta b(l)$, for separations $l = 2\delta x$ (black
lines), $10\delta x$ (cyan lines), and $100\delta x$ (magenta lines)
for (a) ${\rm Pr_M}=0.1$, (b) ${\rm Pr_M}=1$, and (c) ${\rm Pr_M}=10$
(cf. Figs. $23$ (b.3), (c.3), and (d.3), for 3DMHD, in
Ref.~\cite{sahoomhd}).  Arguments of these PDFs are scaled by their
root-mean-square values.}
\label{fig:incrementpdfs}
\end{figure}

Figures\ref{fig:incrementpdfs} (a), (b), and (c) show these PDFs for ${\rm
Pr_M}=0.1$, ${\rm Pr_M}=1$, and ${\rm Pr_M}=10$, respectively. The PDFs of
velocity increments are shown for separations $l = 2\delta x$ (red lines),
$l=10\delta x$ (green lines), and $l=100\delta x$ (blue lines), where $\delta
x$ is our real-space lattice spacing; for PDFs of magnetic-field increments we
also use the separations $l = 2\delta x$ (black lines), $l = 10\delta x$ (cyan
lines), and $l = 100\delta x$ (magenta lines). As in 3DMHD (cf. Figs. $23$
(b.3), (c.3), and (d.3) in Ref.~\cite{sahoomhd}), we see that these
PDFs are nearly Gaussian if the length scale $l$ is large. As $l$ decreases,
the PDFs develop long, non-Gaussian tails, a clear signature of intermittency.
Furthermore, a comparison of the red and black lines in these plots indicates
that the PDFs of the magnetic-field increments are broader than their velocity
counterparts for ${\rm Pr_M}=1$ and ${\rm Pr_M}=10$ and the trend reverses for
${\rm Pr_M}=0.1$.  This is again because of the stronger intermittency in the
magnetic-field than that in velocity field at large ${\rm Pr_M}$, a result we
have also found for 3DMHD~\cite{sahoomhd,sahoophd} from DNS studies with higher
than those presented for 3DRFMHD here. 

The general trend that we notice from these PDFs is that the magnetic field
is more intermittent than the velocity field at large Prandtl numbers but
this difference decreases as ${\rm Pr_M}$ is lowered. We quantify this below
by obtaining multiscaling exponents of velocity and magnetic-field structure
functions in subsection~\ref{sec:stfn}.

\subsubsection{PDFs of angles}

The angles between various vectors, in 3DRFMHD turbulence, give us some
understanding of the nature of flows. Here we present PDFs of cosines of
angles between various vectors and the three eigenvectors of ${\mathbb S}$
that give directions of stretching and compression.

\begin{figure}[htb]
\begin{center}
\includegraphics[width=0.95\textwidth]{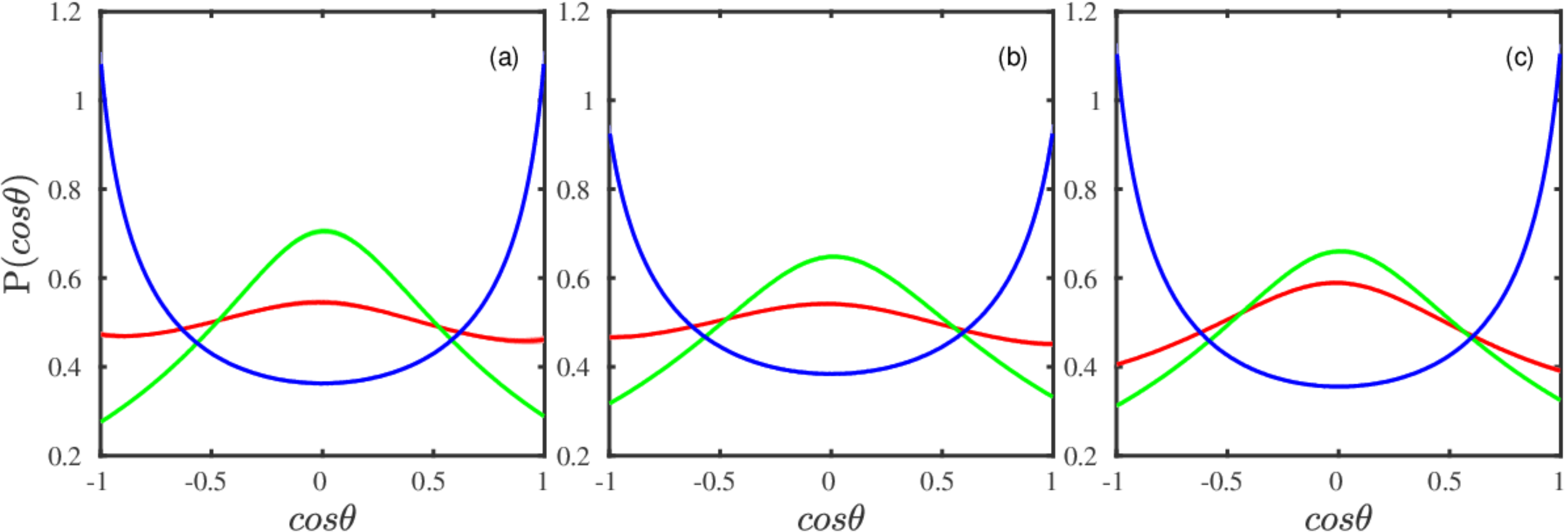}
\end{center}
\caption[]{Plots of the PDFs of cosines of angles between
the vorticity $\bomega$ and the eigenvectors of the fluid rate-of-strain
tensor ${\mathbb S}$, namely $\hat e^1_u$ [red line], $\hat e^2_u$
[blue line], and $\hat  e^3_u$ [green line]; for (a) ${\rm
Pr_M}=0.1$, (b) ${\rm Pr_M}=1$, and (c) ${\rm Pr_M}=10$
(cf. Figs. $13$ for \textit{decaying} 3DMHD turbulence in Ref.~\cite{sahoomhd}). One-standard-deviation error bars are indicated by the shaded regions.}
\label{fig:angle-w-eu}
\end{figure}
\begin{figure}[htb]
\begin{center}
\includegraphics[width=0.95\textwidth]{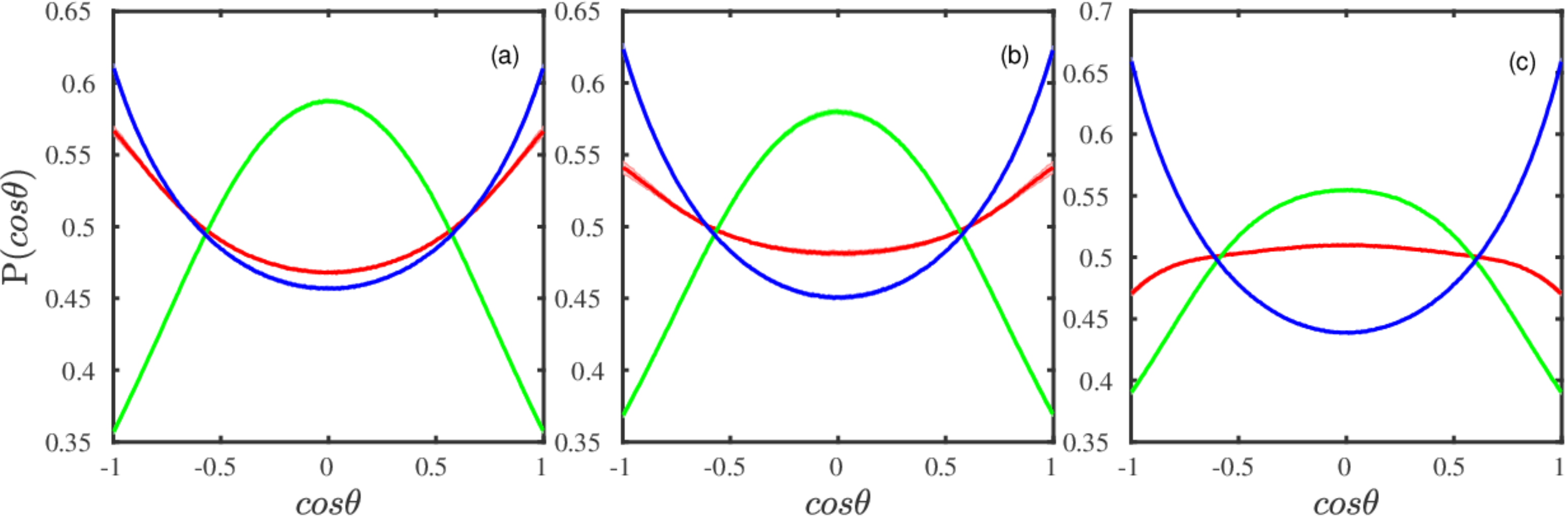}
\end{center}
\caption[]{Plots of the PDFs of cosines of angles between
the current density $\bfj$ and the eigenvectors of the fluid rate-of-strain
tensor ${\mathbb S}$, namely $\hat e^1_u$ [red line], $\hat e^2_u$
[blue line], and $\hat  e^3_u$ [green line]; for (a) ${\rm
Pr_M}=0.1$, (b) ${\rm Pr_M}=1$, and (c) ${\rm Pr_M}=10$ (cf. Fig. $14$
for \textit{decaying} 3DMHD turbulence in Ref.~\cite{sahoomhd}). 
One-standard-deviation error bars are indicated by the shaded regions.}
\label{fig:angle-j-eu}
\end{figure}
\begin{figure}[htb]
\begin{center}
\includegraphics[width=0.95\textwidth]{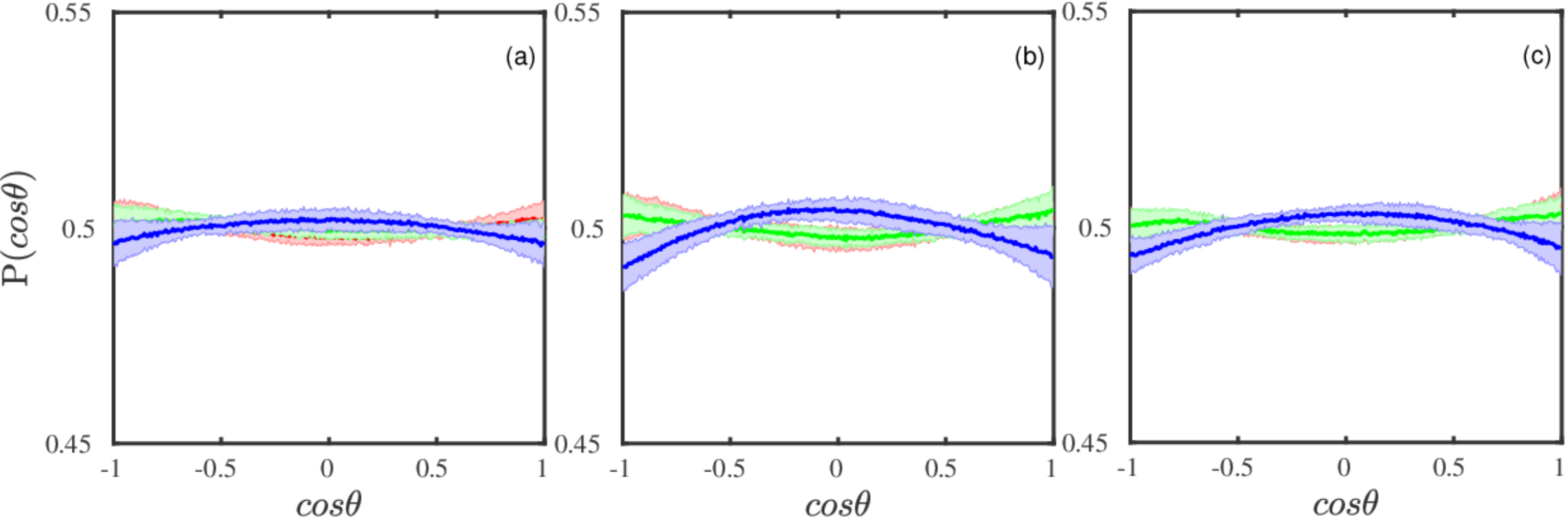}
\end{center}
\caption[]{Semilog (base 10) plots of the PDFs of cosines of angles between
the velocity $\bfu$ and the eigenvectors of the fluid rate-of-strain tensor
${\mathbb S}$, namely $\hat e^1_u$ [red line], $\hat e^2_u$ [blue line], and $\hat  e^3_u$ [green line]; for (a) ${\rm Pr_M}=0.1$, (b) ${\rm Pr_M}=1$, and (c) ${\rm Pr_M}=10$
(cf. Fig. $15$ for \textit{decaying} 3DMHD turbulence in Ref.~\cite{sahoomhd}). 
One-standard-deviation error bars are indicated by the shaded regions.}
\label{fig:angle-u-eu}
\end{figure}
\begin{figure}[htb]
\begin{center}
\includegraphics[width=0.95\textwidth]{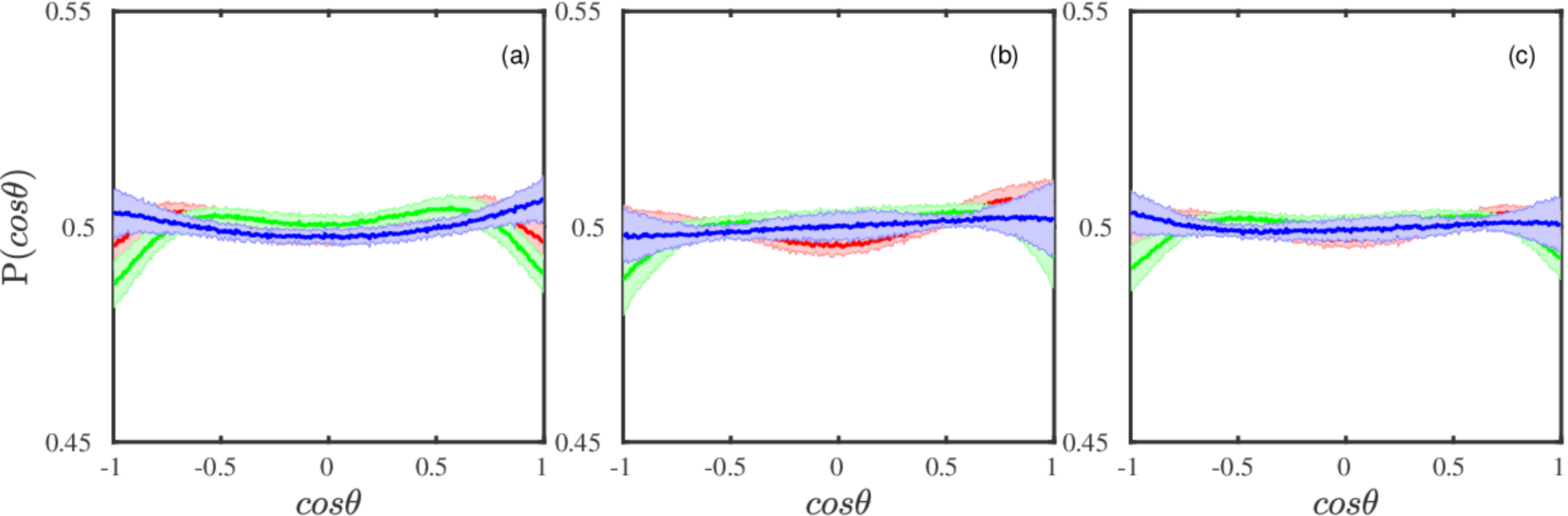}
\end{center}
\caption[]{Plots of the PDFs of cosines of angles between
the magnetic field $\bfb$ and the eigenvectors of the fluid rate-of-strain
tensor ${\mathbb S}$, namely $\hat e^1_u$ [red line], $\hat e^2_u$
[blue line], and $\hat  e^3_u$ [green line]; for (a) ${\rm
Pr_M}=0.1$, (b) ${\rm Pr_M}=1$, and (c) ${\rm Pr_M}=10$
(cf. Fig. $16$ for \textit{decaying} 3DMHD turbulence in Ref.~\cite{sahoomhd}). 
One-standard-deviation error bars are indicated by the shaded regions.}
\label{fig:angle-b-eu}
\end{figure}
\begin{figure}[htb]
\begin{center}
\includegraphics[width=0.95\textwidth]{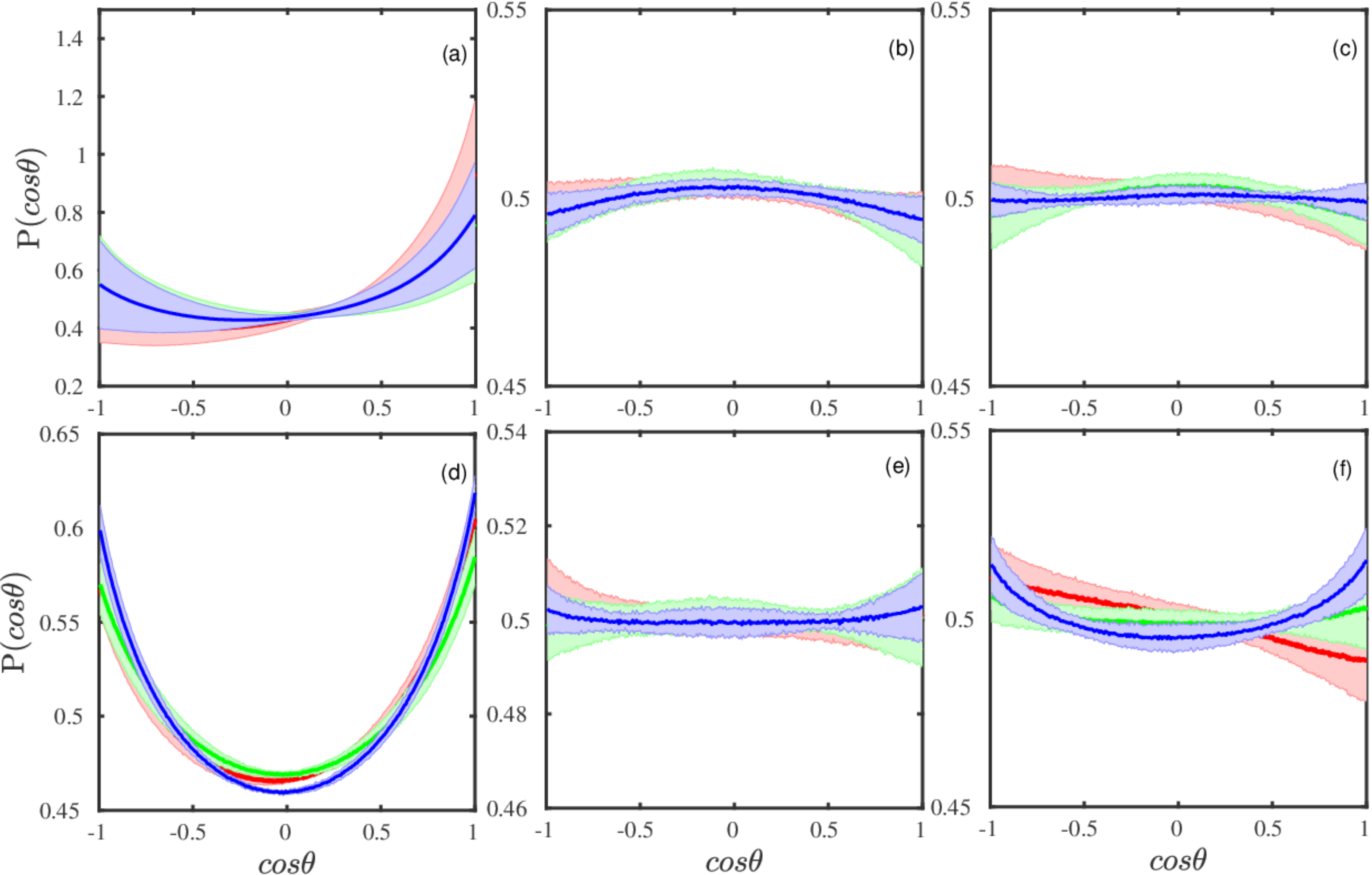}\\
\end{center}
\caption[]{Plots of PDFs of cosines of angles
between (a) $\bfu$ and $\bfb$, (b) $\bfu$ and $\bomega$, 
(c) $\bfu$ and $\bfj$, (d) $\bomega$ and $\bfj$, (e) $\bfb$ and $\bomega$,
and (f) $\bfb$ and $\bfj$ for ${\rm Pr_M}=0.1$ (green line),
${\rm Pr_M}=1$ (red line), ${\rm Pr_M}=10$ (blue line)
(cf. Fig. $17$ for \textit{decaying} 3DMHD turbulence in Ref.~\cite{sahoomhd}). 
One-standard-deviation error bars are indicated by the shaded regions.}
\label{fig:angle-pdf-all}
\end{figure} 

In figures~\ref{fig:angle-w-eu} (a), (b), and (c) we show, for ${\rm Pr_M}=0.1$,
${\rm Pr_M}=1$, and ${\rm Pr_M}=10$, respectively, the PDFs of cosines of
angles between the vorticity $\bomega$ and the eigenvectors of the fluid
rate-of-strain tensor ${\mathbb S}$, namely $\hat e^1_u$ [red line], $\hat
e^2_u$ [blue line], and $\hat  e^3_u$ [green line]. As in 3DMHD (cf. Figs. $13$
for \textit{decaying} 3DMHD turbulence in Ref.~\cite{sahoomhd}), we find that
$\bomega$ is preferentially aligned or anti-aligned with $\hat e^2_u$
for all ${\rm Pr_M}$ in 3DRFMHD. There is some tendency for perpendicular
alignment of $\bomega$ with $\hat e^1_u$ and $\hat e^3_u$ for all ${\rm Pr_M}$
and no preferential angle with $\hat e^3_u$.

In figures~\ref{fig:angle-j-eu} (a), (b), and (c) we show, for ${\rm Pr_M}=0.1$,
${\rm Pr_M}=1$, and ${\rm Pr_M}=10$, respectively, the PDFs of cosines of
angles between the current density $\bfj$ and the eigenvectors of the fluid
rate-of-strain tensor ${\mathbb S}$, namely $\hat e^1_u$ [red line], $\hat
e^2_u$ [blue line], and $\hat  e^3_u$ [green line]. The alignment or
anti-alignment tendencies in this case are not as strong as for $\bomega$. We
see that $\bfj$ has some tendency to be perpendicular to $\hat e^3_u$ and shows
preferential alignment or anti-alignment along $\hat e^1_u$ and $\hat e^2_u$
for all values of ${\rm Pr_M}$ we have studied. These PDFs for 3DRFMHD are
slightly different from their counterparts in 3DMHD~\cite{sahoomhd,sahoophd}; but 
we note that Fig. $14$ in Ref.~\cite{sahoomhd} has data only
for \textit{decaying} 3DMHD turbulence.

In figures~\ref{fig:angle-u-eu} (a), (b), and (c), we show  for ${\rm Pr_M}=0.1$,
${\rm Pr_M}=1$, and ${\rm Pr_M}=10$, respectively, PDFs of the cosines of the
angles between the velocity field $\bfu$ and the eigenvectors of the fluid
rate-of-strain tensor ${\mathbb S}$, namely $\hat e^1_u$ [red line], $\hat
e^2_u$ [blue line], and $\hat  e^3_u$ [green line]. These PDFs indicate that
there is no preferred angle between the velocity field and the eigenvectors of
${\mathbb S}$ for all the values of ${\rm Pr_M}$ we have studied. This
behavior of 3DRFMHD is completely different from that found for decaying
3DMHD~\cite{sahoomhd,sahoophd} turbulence, where the velocity $\bfu$ tends to
align or anti-align with ${\hat e}_u^2$ and there is a weak tendency for an
angle close to  $45^\circ$ (or $135^\circ$)  between $\bfu$ and ${\hat e}_u^1$
and ${\hat e}_u^3$. We note that Fig. $15$ in Ref.~\cite{sahoomhd} has data
only for \textit{decaying} 3DMHD turbulence.

Figures~\ref{fig:angle-b-eu} (a), (b), and (c) show, for ${\rm Pr_M}=0.1$,
${\rm Pr_M}=1$, and ${\rm Pr_M}=10$, respectively, the PDFs of the cosines of
angles between the magnetic field $\bfb$ and the eigenvectors of the fluid
rate-of-strain tensor ${\mathbb S}$, namely $\hat e^1_u$ [red line], $\hat
e^2_u$ [blue line], and $\hat  e^3_u$ [green line]. At least at the level of
resolution of our calculation, these PDFs do not show any clear trends with
${\rm Pr_M}$. We note that Fig. $16$ in Ref.~\cite{sahoomhd} has data only for
\textit{decaying} 3DMHD turbulence.

Figures~\ref{fig:angle-pdf-all} (a), (b), (c), (d), (e), and (f) show,
respectively, PDFs of the cosines of angles between $\bfu$ and $\bfb$, $\bfu$
and $\bomega$, $\bfu$ and $\bfj$, $\bomega$ and $\bfj$, $\bfb$ and $\bomega$,
and $\bfb$ and $\bfj$ for ${\rm Pr_M}=0.1$ (green line), ${\rm Pr_M}=1$ (red
line), and ${\rm Pr_M}=10$ (blue line). Figure~\ref{fig:angle-pdf-all} (a)
shows that $\bfu$ and $\bfb$ are preferentially aligned for ${\rm Pr_M}=1$ and
${\rm Pr_M}=10$. The alignment or anti-alignment, i.e., the probability of
occurrence of \textit{Beltrami correlations} (cos$\theta = \pm 1$)~\cite{dmit}
implies the depletion of nonlinearity. From figures~\ref{fig:angle-pdf-all}
(b) and (c) we see that there is no preferred angle between $\bfu$
and $\bomega$ and between $\bfu$ and $\bfj$; and there is very little
Prandtl-number dependence in the PDFs shown in these figures.  In
Fig.~\ref{fig:angle-pdf-all} (d), we see $\bomega$ and $\bfj$ are
preferentially aligned or anti-aligned for all ${\rm Pr_M}$. From
Figure~\ref{fig:angle-pdf-all} (e), we see there is no particularly  preferred
angle between $\bfb$ and $\bomega$ for ${\rm Pr_M}=0.1$ , ${\rm Pr_M}=1$ and
${\rm Pr_M}=10$.  Figure~\ref{fig:angle-pdf-all} (f) shows preferential
alignment or anti-alignment between $\bfb$ and $\bfj$, with a marginally
greater tendency for the former than the latter, for ${\rm Pr_M} = 1$.  Figure
$17$ in Ref.~\cite{sahoomhd} has data only for \textit{decaying} 3DMHD
turbulence.


\subsection{Structure functions \label{sec:stfn}}

We continue our elucidation of intermittency in 3DRFMHD turbulence by studying
the scale dependence of order-$p$ equal-time, longitudinal velocity structure
functions $S_p^u(l) \equiv \langle|\delta u_{\parallel}(\bfx,l)|^p\rangle$ and
longitudinal magnetic-field structure functions $S_p^b(l) \equiv \langle|\delta
b_{\parallel}(\bfx,l)|^p\rangle$, respectively, where $\delta
u_{\parallel}(\bfx,l) \equiv
\bfu(\bfx+\bfl,t)-\bfu(\bfx,t)]\cdot\frac{\bfl}{l}$ and $\delta
b_{\parallel}(\bfx,l) \equiv
\bfb(\bfx+\bfl,t)-\bfb(\bfx,t)]\cdot\frac{\bfl}{l}$. From these structure
functions we also obtain the hyperflatnesses $F_6^u(l)=S_6^u(l)/[S_2^u(l)]^3$
and $F_6^b(l)=S_6^b(l)/[S_2^b(l)]^3$.  

\begin{figure}[htb]
\begin{center}
\includegraphics[width=0.95\textwidth]{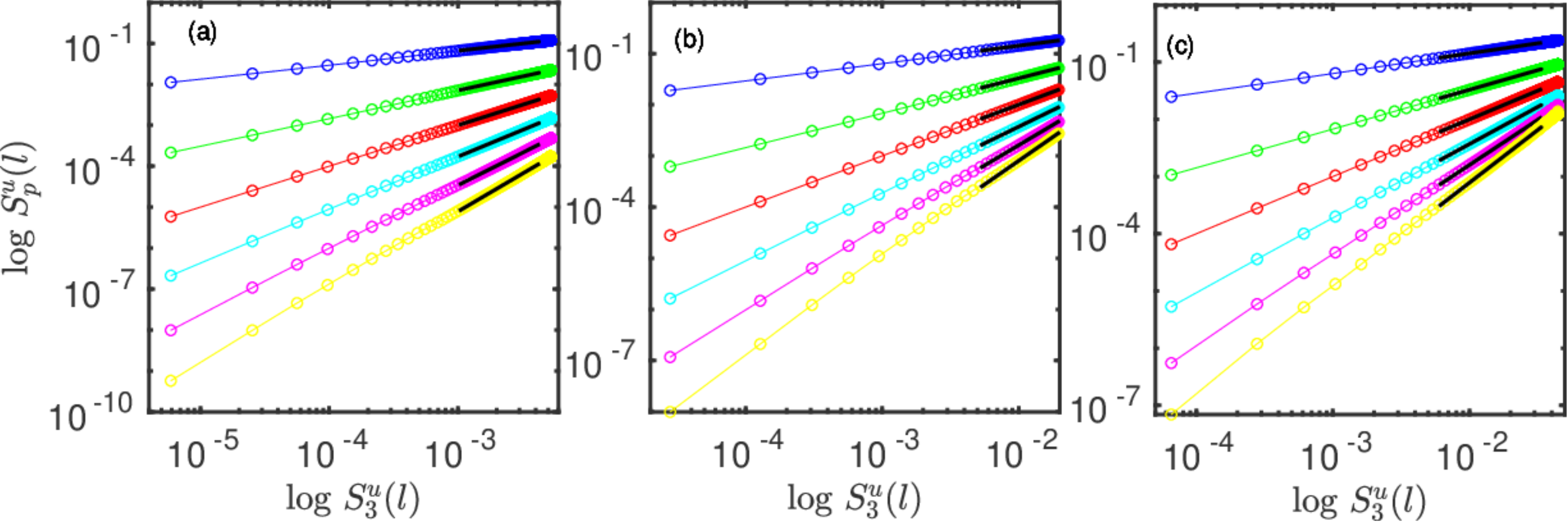}
\end{center}
\caption[]{Log-log (base 10) extended-self-similarity (ESS) plots of 
velocity structure functions of
order $p$ versus that of order $3$; $p=1$ (blue), $p=2$ (green), $p=3$ (red),
$p=4$ (cyan), $p=5$ (magenta), and $p=6$ (yellow); for (a) ${\rm Pr_M}=0.1$,
(b) ${\rm Pr_M}=1$, and (c) ${\rm Pr_M}=10$. The region marked with black
line shows the range over which we calculate the exponent ratio
$\zeta^u_p/\zeta^u_3$ (cf. Figs. $25$ (b.1), (c.1), and (d.1) in Ref.~\cite{sahoomhd})}.
\label{fig:stfn-u}
\end{figure}
\begin{figure}[htb]
\begin{center}
\includegraphics[width=0.95\textwidth]{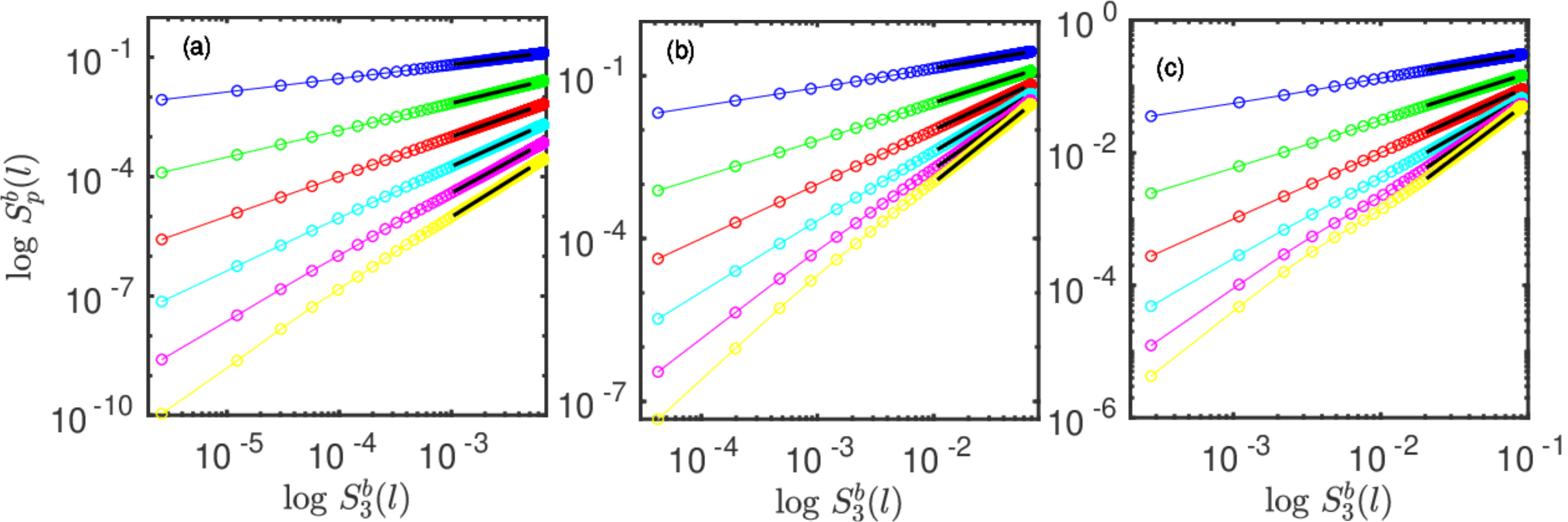}
\end{center}
\caption[]{Log-log (base 10) ESS plots of magnetic-field structure functions
of order $p$ versus that of order $3$; $p=1$ (blue), $p=2$ (green), $p=3$
(red), $p=4$ (cyan), $p=5$ (magenta), and $p=6$ (yellow); for (a) ${\rm
Pr_M}=0.1$, (b) ${\rm Pr_M}=1$, and (c) ${\rm Pr_M}=10$. The region marked
with a black line shows the range over which we calculate the exponent ratio
$\zeta^b_p/\zeta^b_3$ (cf. Figs. $25$ (b.2), (c.2), and (d.2) in 
Ref.~\cite{sahoomhd}).}
\label{fig:stfn-b}
\end{figure}
\begin{figure}[htb]
\begin{center}
\includegraphics[width=0.95\textwidth]{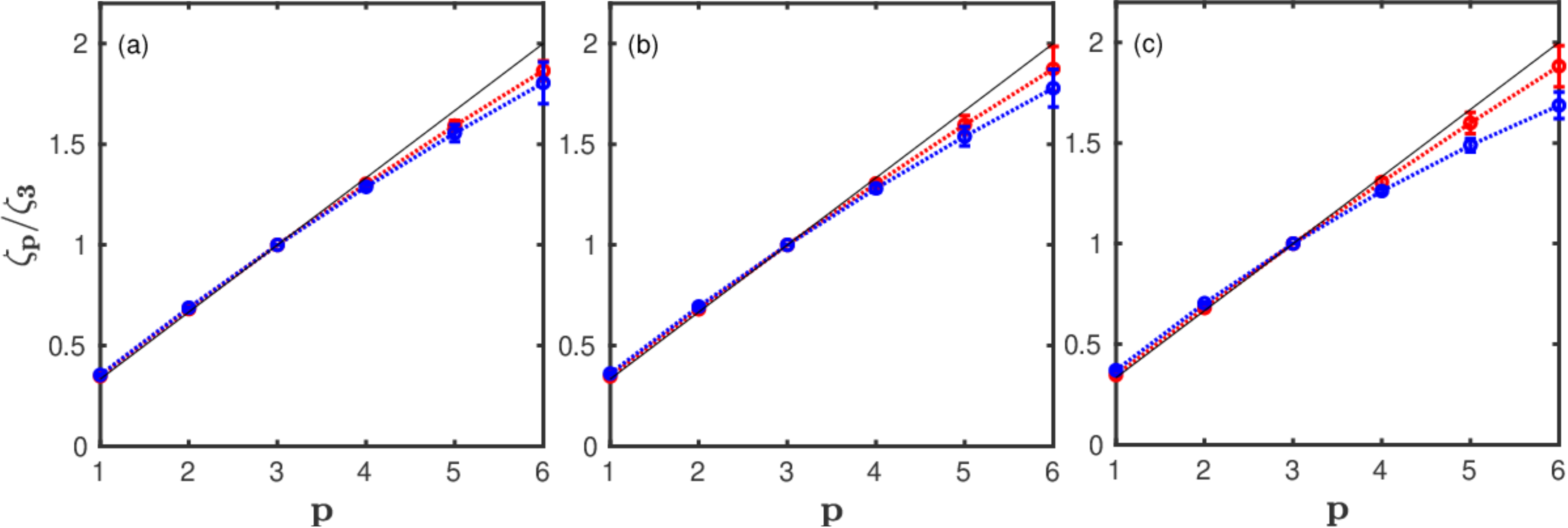}
\end{center}
\caption[]{Multiscaling exponent ratios $\zeta^u_p/\zeta^u_3$ (red line) and
$\zeta^b_p/\zeta^b_3$ (blue line) versus $p$ for (a) ${\rm Pr_M}=0.1$, (b)
	${\rm Pr_M}=1$, and (c) ${\rm Pr_M}=10$ (cf. Figs. $25$ (b.3), (c.3), and (d.3) in Ref.~\cite{sahoomhd}).}
\label{fig:zetap}
\end{figure}
\begin{figure}[htb]
\begin{center}
\includegraphics[width=0.95\textwidth]{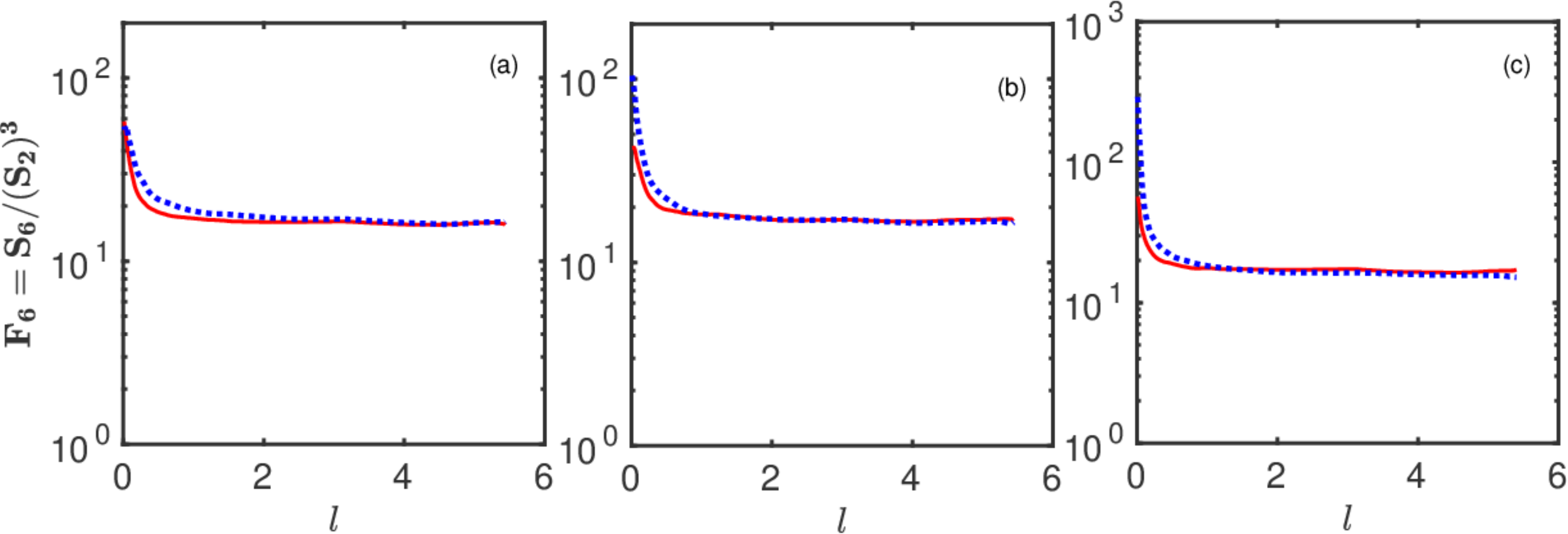}
\end{center}
\caption[]{Semilog (base 10) plots of the hyperflatness of the velocity field
$F^u_6(l)=\frac{S^u_6(l)}{|S^u_3(l)|^3}$ (red line) and its magnetic
counterpart $F^b_6(l)=\frac{S^b_6(l)}{|S^b_3(l)|^3}$ (blue dashed line) for
(a) ${\rm Pr_M}=0.1$, (b) ${\rm Pr_M}=1$, and (c) ${\rm Pr_M}=10$ (cf. Figs. 
$25$ (b.4), (c.4), and (d.4) in Ref.~\cite{sahoomhd}).}
\label{fig:f6}
\end{figure}
\begin{table}
\caption{Multiscaling exponent ratios $\zeta_p^u/\zeta_3^u$ 
and $\zeta_p^b/\zeta_3^b$ from our DNS for 3DRFMHD turbulence.}
\label{table:zetap}
\begin{center}
\begin{tabular}{l|l|l|l}
\hline\noalign{\smallskip}
$p$ & $\zeta_p^u/\zeta_3^u$; $\zeta_p^b/\zeta_3^b ({\rm Pr_M}=0.1)$ &
$\zeta_p^u/\zeta_3^u$; $\zeta_p^b/\zeta_3^b ({\rm Pr_M}=1)$ &
$\zeta_p^u/\zeta_3^u$; $\zeta_p^b/\zeta_3^b ({\rm Pr_M}=10)$ \\  
\noalign{\smallskip}\hline\noalign{\smallskip}
1 & $0.35\pm0.00$; $0.35\pm0.00$ & $0.35\pm0.00$;
$0.36\pm0.00$ & $0.35\pm0.01$; $0.37\pm0.00$ \\ 
2 & $0.68\pm0.00$; $0.69\pm0.00$ & $0.68\pm0.00$;
$0.69\pm0.00$ & $0.68\pm0.01$; $0.70\pm0.00$ \\ 
3 & $1.00\pm0.00$; $1.00\pm0.00$ & $1.00\pm0.00$;
$1.00\pm0.00$ & $1.00\pm0.00$; $1.00\pm0.00$ \\ 
4 & $1.30\pm0.01$; $1.29\pm0.01$ & $1.30\pm0.01$;
$1.28\pm0.01$ & $1.30\pm0.02$; $1.26\pm0.01$ \\ 
5 & $1.59\pm0.03$; $1.56\pm0.04$ & $1.60\pm0.05$;
$1.53\pm0.05$ & $1.60\pm0.05$; $1.49\pm0.03$ \\ 
6 & $1.86\pm0.05$; $1.80\pm0.10$ & $1.87\pm0.11$;
$1.78\pm0.10$ & $1.88\pm0.10$; $1.69\pm0.07$ \\ 
\noalign{\smallskip}\hline
\end{tabular}
\end{center}
\end{table}


For the inertial range $\eta_d^u, \eta_d^b \ll l \ll L$, we expect
$S_p^u(l)\sim l^{\zeta_p^u}$ and $S_p^b(l)\sim l^{\zeta_p^b}$, where
$\zeta_p^u$ and $\zeta_p^b$ are inertial-range multiscaling exponents for
velocity and magnetic fields, respectively; if these fields show
multiscaling, we expect significant deviations from the K41 result
$\zeta_p^{uK41} = \zeta_p^{bK41} = p/3$.  [We do not have any mean magnetic
field in our simulations, so we do not expect any
Iroshnikov-Kraichnan~\cite{ik} scaling.] Given large inertial ranges, the
multiscaling exponents can be extracted from slopes of log-log plots of
structure functions versus $l$.  In DNSs, such as the ones
presented here, inertial ranges are limited by the spatial resolution, so we use the
extended-self-similarity (ESS) procedure~\cite{benzi93,chakraborty10} in which we
determine the multiscaling exponent ratios $\zeta_p^u/\zeta_3^u$ and
$\zeta_p^b/\zeta_3^b$, respectively, from slopes of log-log plots of $S_p^u$
versus $S_3^u$ and $S_p^b$ versus $S_3^b$. The structure functions with ESS
are shown in figures~\ref{fig:stfn-u} and \ref{fig:stfn-b}.

Figures~\ref{fig:zetap} show the plots of exponent ratios
$\zeta_p^u/\zeta_3^u$ (red line with circles) and $\zeta_p^b/\zeta_3^b$ (blue
line with asterisks) versus order $p$ for (a) ${\rm Pr_M}=0.1$, (b) ${\rm
Pr_M}=1$, and (c) ${\rm Pr_M}=10$. The black solid line shows the K41 result
for comparison. These plots show significant deviations from the K41 result,
especially for $p > 3$, as in 3DMHD~\cite{sahoomhd,sahoophd}, both for velocity and
magnetic fields; this provides an effective measure of inertial-range
intermittency. We observe from these plots that, at large values of ${\rm
Pr_M}$, the magnetic field is more intermittent than the velocity field as
the deviations of $\zeta_p^b/\zeta_3^b$ from the K41 result $p/3$ are larger
than those of $\zeta_p^u/\zeta_3^u$. This behavior in 3DRFMHD is similar to its 
counterpart in 3DMHD ~\cite{sahoomhd,sahoophd} (cf. Figs. $25$ (b.3), (c.3), and 
(d.3) in Ref.~\cite{sahoomhd}).

Intermittency is also apparent in plots versus $l$ of the hyperflatnesses
$F^u_6(l)=\frac{S^u_6(l)}{S^u_2(l)^3}$ (red line) and
$F^b_6(l)=\frac{S^b_6(l)}{S^b_2(l)^3}$ (blue dashed line) in
Figs.~\ref{fig:f6} for (a) ${\rm Pr_M}=0.1$, (b) ${\rm Pr_M}=1$, and (c)
${\rm Pr_M}=10$. As $l$ decreases, $F^b_6(l)$ rises more rapidly than
$F^u_6(l)$ except for ${\rm Pr_M} = 0.1$, in consonance with the trends
mentioned in the previous paragraph.  

\subsection{Isosurfaces in RFMHD \label{sec:isosurface}}

Isosurfaces of moduli of the vorticity and current density, $\omega$ and $j$,
respectively, or energy dissipation rates and the effective pressure are
useful ways of visualising structures in 3DMHD flows. In fluid turbulence
isosurfaces of $\omega$ show slender, tube-like structures if $\omega$ is
large; by contrast, in 3DMHD turbulence isosurfaces of $\omega$ and $j$ are
organized into sheets~\cite{sahoomhd,sahoophd} (cf. Figs. $26$ (b.3), (c.3), (d.3), 
and $27$ (b.3), (c.3), (d.3) in Ref.~\cite{sahoomhd}). Isosurfaces of the
energy-dissipation-rates show sheets in both fluid and 3DMHD turbulence; and
isosurfaces of the pressure in fluid turbulence, and of the effective
pressure in 3DMHD turbulence are like clouds, with some tube-type structures
in fluid turbulence.

\begin{figure}[htb]
\begin{center}
\includegraphics[width=0.95\textwidth]{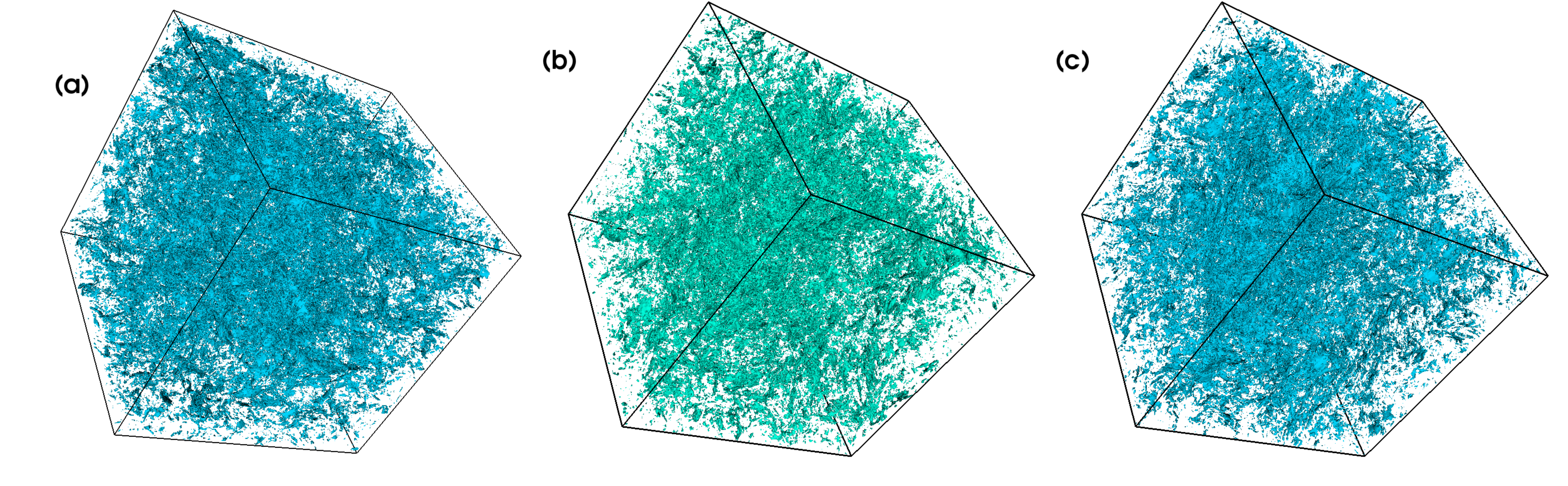}
\end{center}
\caption[]{Isosurfaces of the modulus of the vorticity $\omega$ in 3DRFMHD
for (a) ${\rm Pr_M}=0.1$, (b) ${\rm Pr_M}=1$, and (c) ${\rm Pr_M}=10$ and
values of $\omega$ that are two standard deviations more than its mean 
	value (cf. Figs. $26$ (b.3), (c.3), and (d.3) in Ref.~\cite{sahoomhd}).} 
\label{fig:iso-omega}
\end{figure}
\begin{figure}[htb]
\begin{center}
\includegraphics[width=0.95\textwidth]{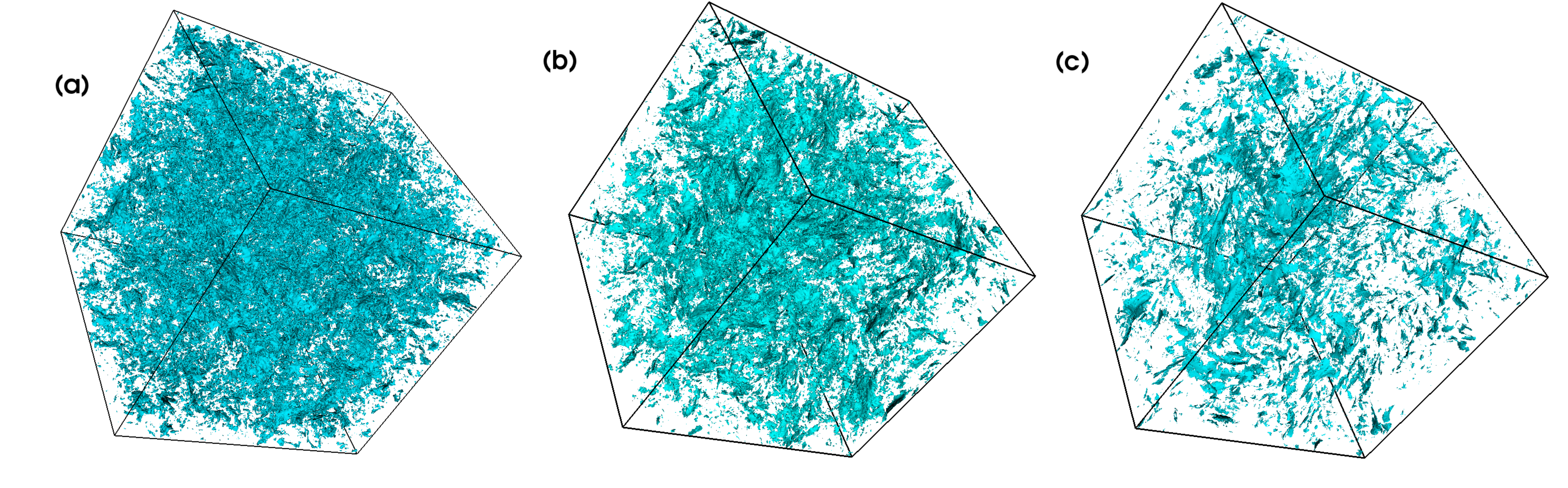}
\end{center}
\caption[]{Isosurfaces of the modulus of the current density $j$ in 3DRFMHD
for (a) ${\rm Pr_M}=0.1$, (b) ${\rm Pr_M}=1$, and (c) ${\rm Pr_M}=10$ and
values of $j$ that are two standard deviations more than its mean
	value (cf. Figs. $27$ (b.3), (c.3), and (d.3) in Ref.~\cite{sahoomhd}).} 
\label{fig:iso-j}
\end{figure}
\begin{figure}[htb]
\begin{center}
\includegraphics[width=0.95\textwidth]{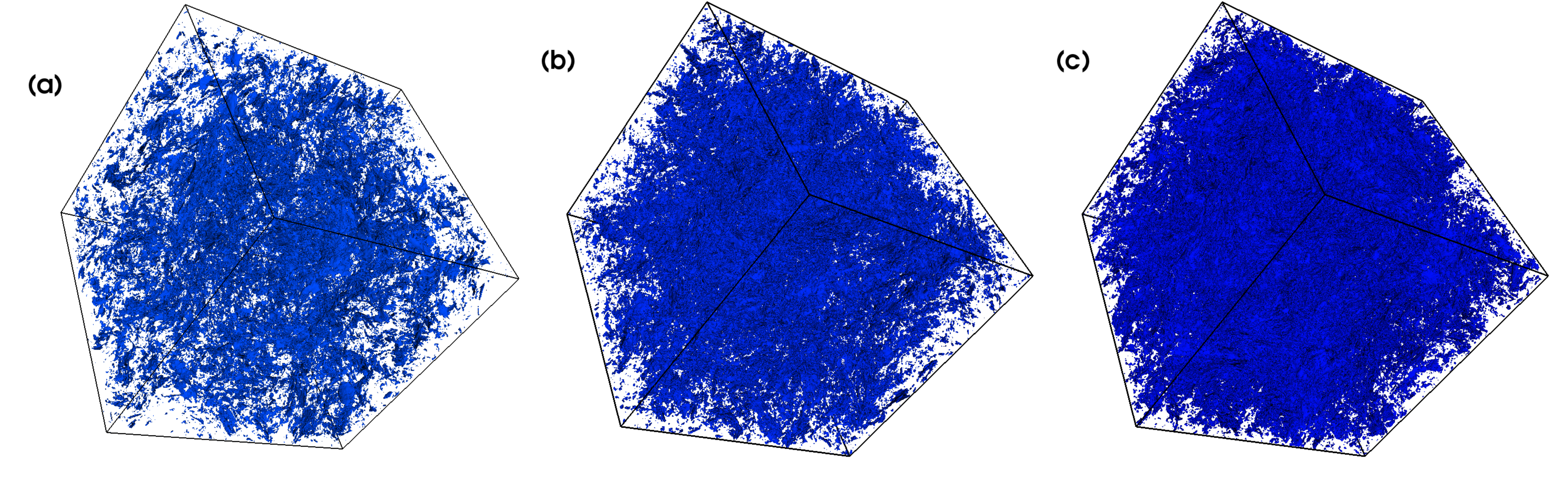}
\end{center}
\caption[]{Isosurfaces of $\epsilon_u$ in 3DRFMHD for (a) ${\rm Pr_M}=0.1$,
(b) ${\rm Pr_M}=1$, and (c) ${\rm Pr_M}=10$ and values of $\epsilon_u$ that
are two standard deviations more than its mean value (cf. Figs. $28$ (b.3), 
(c.3), and (d.3) in Ref.~\cite{sahoomhd}).} 
\label{fig:iso-epsu}
\end{figure}
\begin{figure}[htb]
\begin{center}
\includegraphics[width=0.95\textwidth]{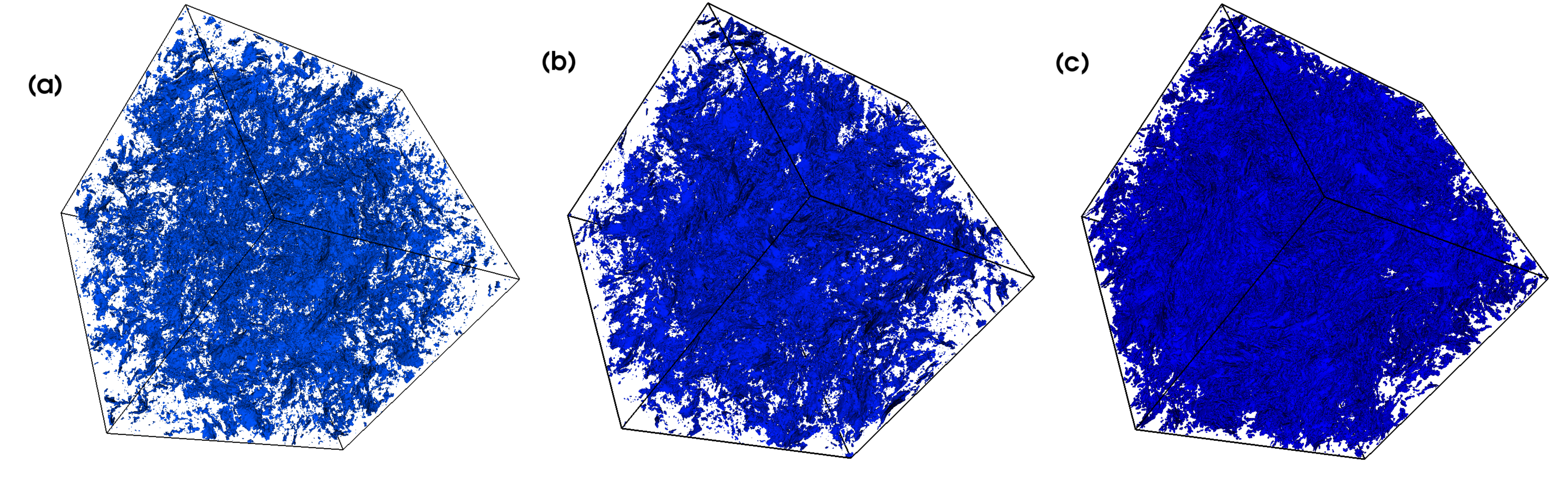}
\end{center}
\caption[]{Isosurfaces of $\epsilon_b$ in 3DRFMHD 
for (a) ${\rm Pr_M}=0.1$, (b) ${\rm Pr_M}=1$, and (c) ${\rm Pr_M}=10$ and
values of $\epsilon_b$ that two standard deviations more than its mean 
value (cf. Figs. $29$ (b.3), (c.3), and (d.3) in Ref.~\cite{sahoomhd})}.
\label{fig:iso-epsb}
\end{figure}
\begin{figure}[htb]
\begin{center}
\includegraphics[width=0.95\textwidth]{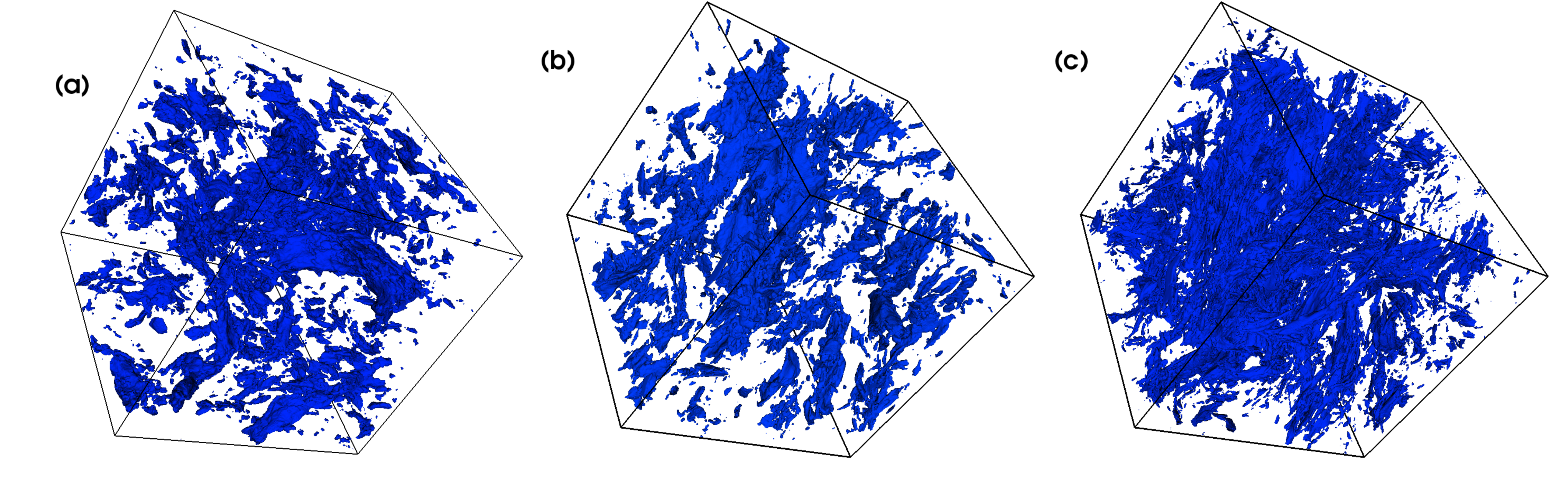}
\end{center}
\caption[]{Isosurfaces of the effective pressure in 3DRFMHD
for (a) ${\rm Pr_M}=0.1$, (b) ${\rm Pr_M}=1$, and (c) ${\rm Pr_M}=10$ and 
values the effective pressure that are two standard deviations more than 
its mean value (cf. Figs. $30$ (b.2), (c.2), and (d.2) in Ref.~\cite{sahoomhd})}.
\label{fig:iso-pressure}
\end{figure}

In Figs.~\ref{fig:iso-omega}, \ref{fig:iso-j}, \ref{fig:iso-epsu},
\ref{fig:iso-epsb}, and \ref{fig:iso-pressure} we show, respectively,
isosurfaces of $\omega$, $j$, the local kinetic-energy-dissipation rate
$\epsilon_u$, the local magnetic-energy-dissipation rate $\epsilon_b$, and the
local effective pressure ${\bar p}$ for 3DRFMHD; the values of these quantities
are chosen to be two standard deviations more than their mean values; the
subplots in these figures show isosurface for (a) ${\rm Pr_M}=0.1$, (b) ${\rm
Pr_M}=1$, and (c) ${\rm Pr_M}=10$. Even a cursory comparison of these
isosurfaces with their analogues for 3DMHD~\cite{sahoomhd,sahoophd} shows that
they are modified dramatically by the random, power-law forcing we employ here:
in particular, isosurfaces of $\omega$, $j$, $\epsilon_u$, $\epsilon_b$, and
the effective pressure appear to be completely stippled versions of the
sheet-type isosurfaces that are seen in 3DMHD
turbulence~\cite{sahoomhd,sahoophd}; this is reminiscent of the destruction of
tube-type isosurfaces of $\omega$ by power-law, random forcing in 3D
Navier-Stokes turbulence~\cite{sain98}.

\subsection{Joint PDFs \label{sec:jpdfs}}

In this subsection we present a set of joint PDFs for 3DRFMHD and
we compare them with their analogues for 3DMHD~\cite{sahoomhd,sahoophd}.
In particular, we give $QR$ plots, and joint PDFs of $\omega$ and
$j$, of $\epsilon_u$ and $\epsilon_b$, of $\omega$ and
$\epsilon_u$, and of $j$ and $\epsilon_b$.

\begin{figure}[htb]
\begin{center}
\includegraphics[width=0.95\textwidth]{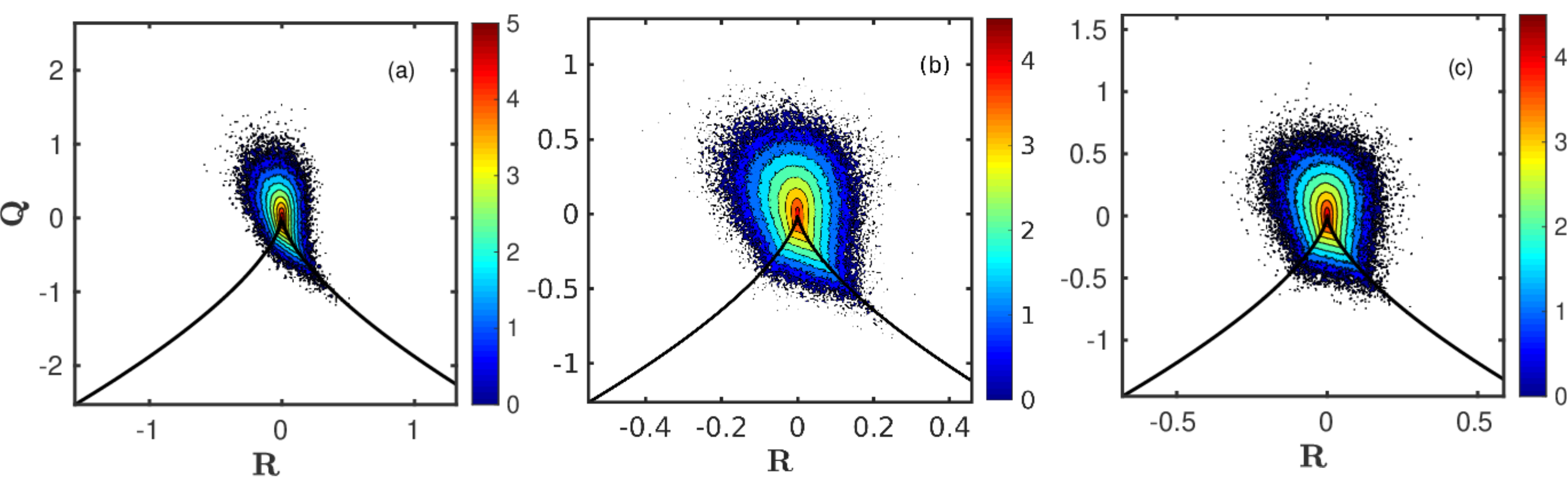}
\end{center}
\caption[]{$QR$ plots, i.e., joint PDFs of $Q$ and $R$, in 3DRFMHD
shown as filled contour plots on a logarithmic scale for (a)
${\rm Pr_M}=0.1$, (b) ${\rm Pr_M}=1$, and (c) ${\rm Pr_M}=10$.
The arguments $Q$ and $R$ of the joint PDF are normalized by
$\langle\omega^2\rangle$ and $\langle\omega^2\rangle^{3/2}$,
respectively. The black curve shows the zero-discriminant line
$D\equiv \frac{27}{4}R^2+Q^3=0$ (cf. Figs. $31$ (b.3), (c.3), and (d.3) in 
Ref.~\cite{sahoomhd}).}

\label{fig:qrplot}
\end{figure}

Figures~\ref{fig:qrplot} (a), (b), and (c) show $QR$ plots, i.e.,
joint PDFs of $Q$ and $R$ for 3DRFMHD, via filled contour plots
on a logarithmic scale, for ${\rm Pr_M}=0.1$, ${\rm Pr_M}=1$, and
${\rm Pr_M}=10$, respectively. The black curve in these plots is
the zero-discriminant line $D\equiv \frac{27}{4}R^2+Q^3=0$. These
$QR$ plots retain, aside from some distortions, the
characteristic tear-drop structure familiar from such plots fluid
turbulence~\cite{pramana09} and 3DMHD~\cite{sahoomhd,sahoophd}.  Notice
that, as we increase ${\rm Pr_M}$, there is a decrease in
the probability of having large values of $Q$ and $R$, i.e.,
regions of large strain or vorticity are suppressed, as we have
also found in 3DMHD (cf. Figs. $31$ (b.2), (c.2), and (d.2) in 
Ref.~\cite{sahoomhd}).

\begin{figure}[htb]
\begin{center}
\includegraphics[width=0.95\textwidth]{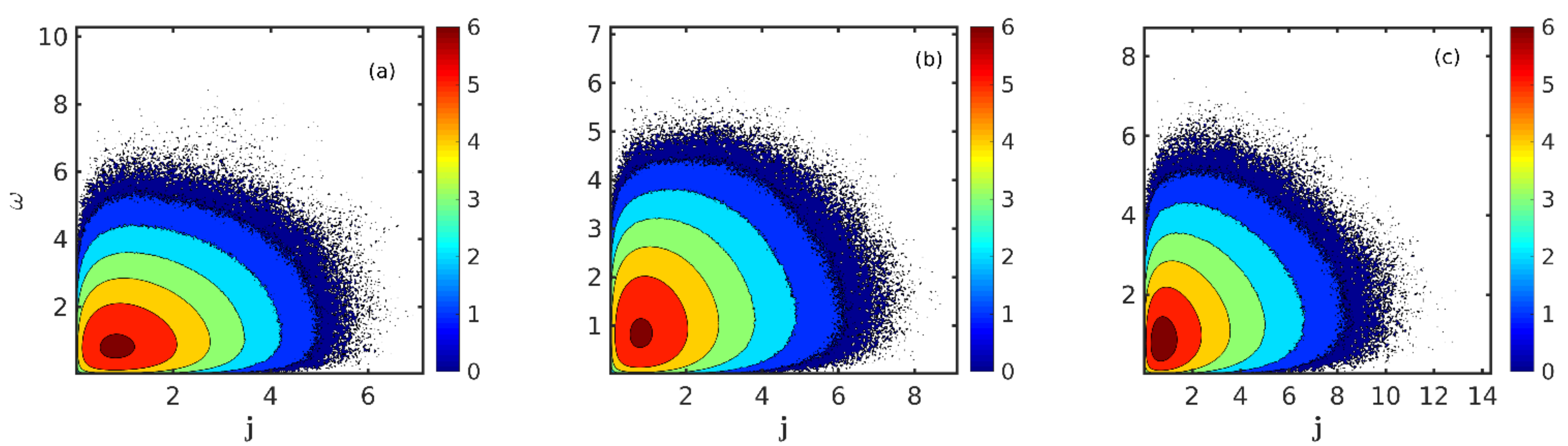}
\end{center}
\caption[]{Joint PDFs of $\omega$ and $j$ for 3DRFMHD shown as
filled contour plots on a logarithmic scale for (a) ${\rm
Pr_M}=0.1$, (b) ${\rm Pr_M}=1$, and (c) ${\rm Pr_M}=10$. The
arguments of the joint PDFs are normalized by their mean values (cf. Figs. $32$ (b.3), 
(c.3), and (d.3) in Ref.~\cite{sahoomhd}).}
\label{fig:jpdf-wj}
\end{figure}

Joint PDFs of $\omega$ and $j$ for ${\rm Pr_M}=0.1$, ${\rm Pr_M}=1$, and ${\rm
Pr_M}=10$ are shown in figures~\ref{fig:jpdf-wj} (a), (b), and (c), respectively,
for 3DRFMHD. These petal-shaped joint PDFs have long tails. As we move away
from ${\rm Pr_M} = 1$ these joint PDFs are distorted such that large values of
$j$ are suppressed if ${\rm Re_{M\lambda}}$ is low, as in
Fig.~\ref{fig:jpdf-wj} (a), whereas large values of $\omega$ are suppressed if
${\rm Re}_{\lambda}$ is low, as in figure~\ref{fig:jpdf-wj} (c).  These trends
are similar to those in 3DMHD turbulence (cf. Figs. $32$ (b.3), (c.3), and
(d.3) in Ref.~\cite{sahoomhd}).

\begin{figure}[htb]
\begin{center}
\includegraphics[width=0.95\textwidth]{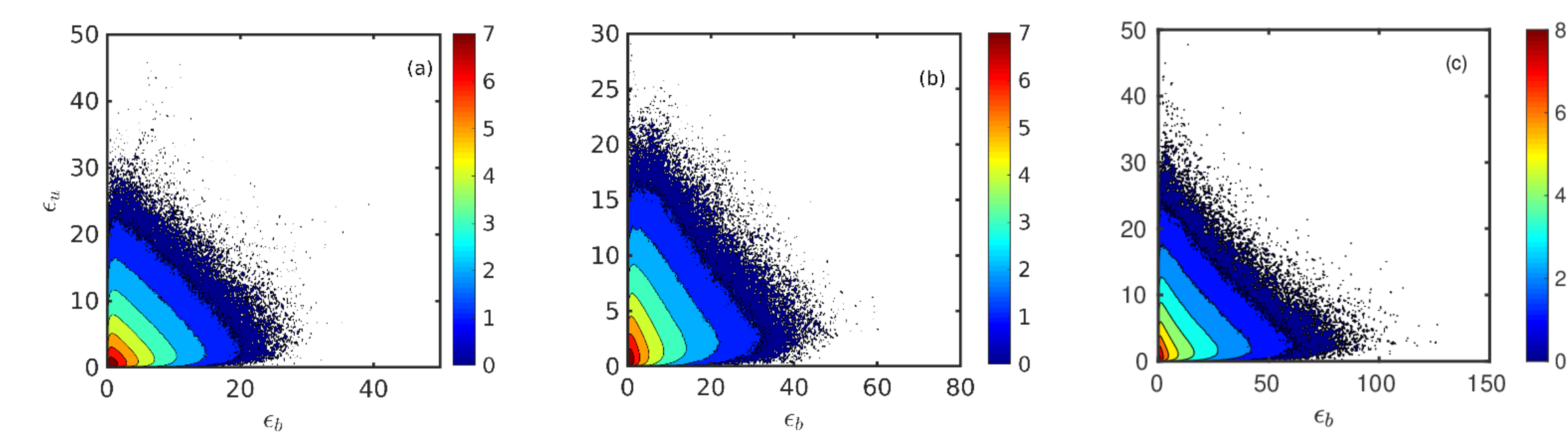}
\end{center}
\caption[]{Joint PDFs of $\epsilon_u$ and $\epsilon_b$ for 3DRFMHD shown as filled
contour plots on a logarithmic scale for (a) ${\rm Pr_M}=0.1$, (b) ${\rm Pr_M}=1$, and
(c) ${\rm Pr_M}=10$. The arguments of the joint PDFs are normalized by their mean
values (cf. Figs. $33$ (b.3), (c.3), and (d.3) in Ref.~\cite{sahoomhd}).}
\label{fig:jpdf-eps}
\end{figure}

Joint PDFs of $\epsilon_u$ and $\epsilon_b$ for ${\rm Pr_M}=0.1$, ${\rm
Pr_M}=1$, and ${\rm Pr_M}=10$ are shown in figures~\ref{fig:jpdf-eps} (a), (b),
and (c).  All these joint PDFs have long tails. As we move away from ${\rm
Pr_M} = 1$ these joint PDFs change slightly such that large values of
$\epsilon_b$ are suppressed if ${\rm Re_{M\lambda}}$ is low, as in
Fig.~\ref{fig:jpdf-eps} (a), whereas large values of $\epsilon_u$ are
suppressed if ${\rm Re}_{\lambda}$ is low, as in Fig.~\ref{fig:jpdf-eps} (c).
These trends are similar to those in 3DMHD turbulence (cf. Figs. $33$ (b.3),
(c.3), and (d.3) in Ref.~\cite{sahoomhd}).

\begin{figure}[htb]
\begin{center}
\includegraphics[width=0.95\textwidth]{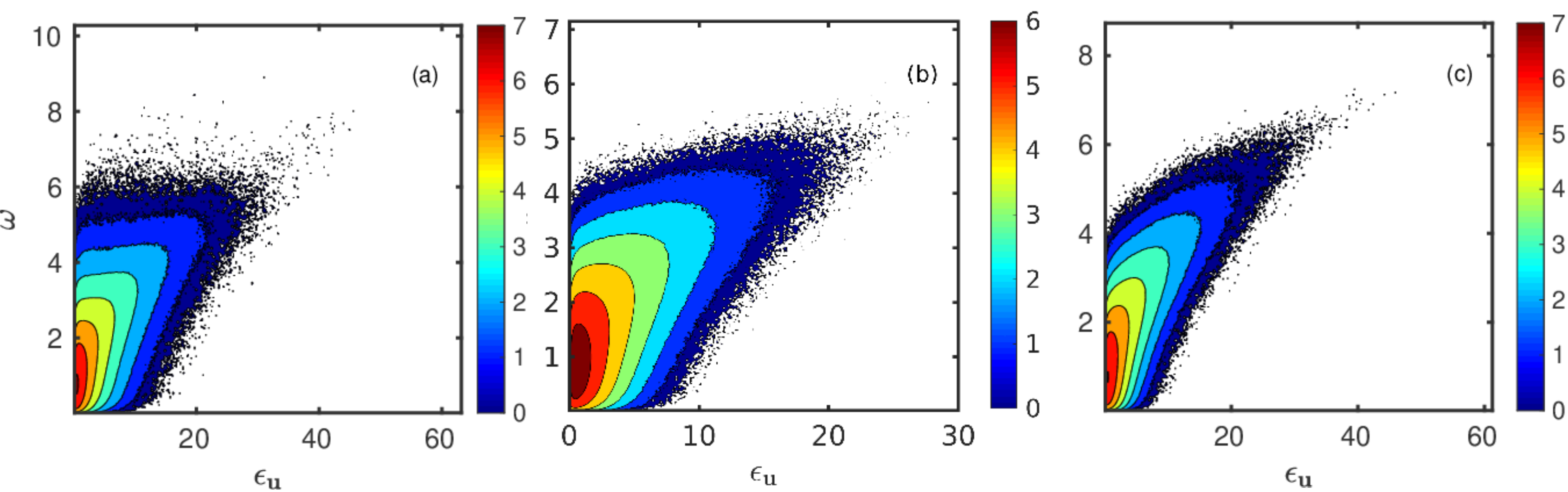}
\end{center}
\caption[]{Joint PDFs of $\omega$ and $\epsilon_u$ for 3DRFMHD shown as filled contour
plots on a logarithmic scale for (a) ${\rm Pr_M}=0.1$, (b) ${\rm Pr_M}=1$, and (c)
${\rm Pr_M}=10$. The arguments of the joint PDFs are normalized by their mean values.}
\label{fig:jpdf-wepsu}
\end{figure}
\begin{figure}[htb]
\begin{center}
\includegraphics[width=0.95\textwidth]{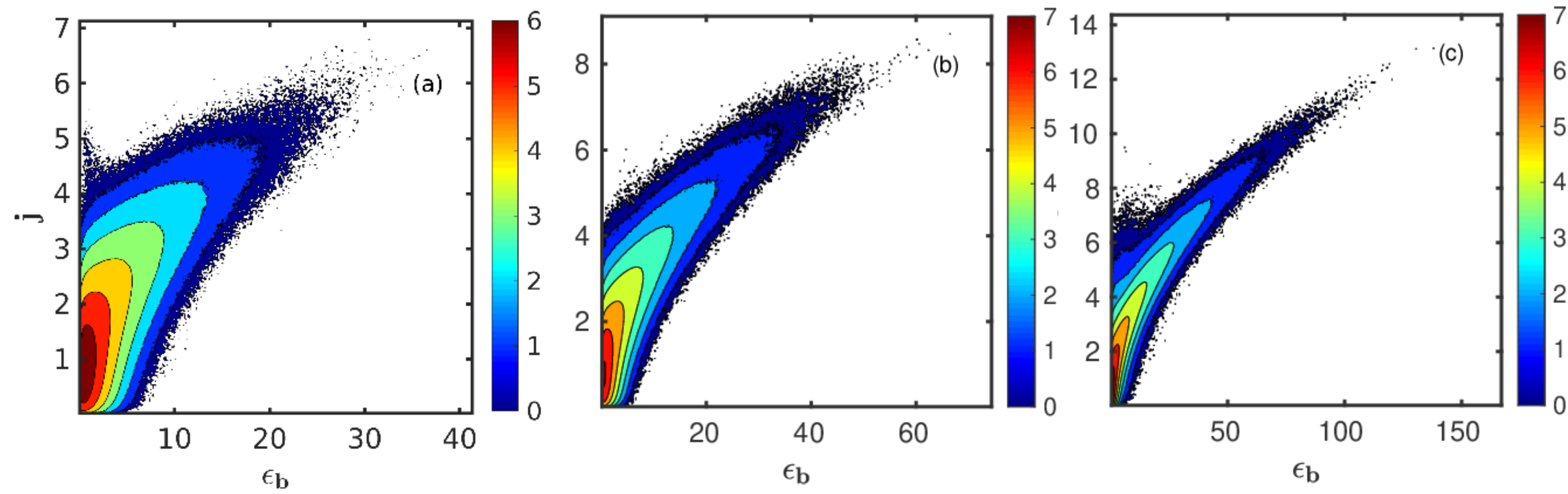}
\end{center}
\caption[]{Joint PDF of $j$ and $\epsilon_b$ for 3DRFMHD shown as filled
contour plots on a logarithmic scale for (a) ${\rm Pr_M}=0.1$, (b) ${\rm Pr_M}=1$, and 
(c) ${\rm Pr_M}=10$. The arguments of the joint PDFs are normalized by their mean values.}
\label{fig:jpdf-jepsb}
\end{figure}

Joint PDFs of $\omega$ and $\epsilon_u$ are shown in Figs.~\ref{fig:jpdf-wepsu}
and those for $j$ and $\epsilon_b$ in figures~\ref{fig:jpdf-jepsb} for (a) ${\rm
Pr_M}=0.1$, (b) ${\rm Pr_M}=1$, and (c) ${\rm Pr_M}=10$ for 3DRFMHD.

\section{Discussions \label{sec:conclusions}}

We have carried out a detailed study of 3DRFMHD turbulence.  Our study has
been designed specifically to study the systematics of the dependence of
these properties on the magnetic Prandtl number ${\rm Pr_M}$ and to compare
them with their counterparts in 3DMHD turbulence.  Our study is restricted to
incompressible MHD turbulence;  we do not include a mean magnetic field as,
e.g., in Refs.~\cite{goldreich95}; furthermore we do not study Lagrangian
properties considered, e.g., in Ref.~\cite{homann07}. In our studies we
obtain (a) various PDFs, such as those of $\omega$, $j$, the energy dissipation rates, 
of the cosines of angles between
various vectors, and scale-dependent velocity and magnetic-field increments,
(b) spectra, e.g., those of the energy and the effective pressure, (c)
velocity and magnetic-field structure functions that can be used to
characterize small-length-scale intermittency, (d) isosurfaces of quantities
such as $\omega$ and $j$, and (e) joint PDFs such as
$QR$ plots.  The evolution of these properties with ${\rm Pr_M}$ has been
described in detail in the previous Section. 

Such a comprehensive study of 3DRFMHD turbulence has not been attempted
heretofore. All earlier studies have concentrated on field-theoretical studies
of correlation functions and spectra in 3DRFMHD; typically these studies use
Fourier-space renormalization group methods, at one- or two-loop levels, or
closures at concentrate on K41-type scaling results~\cite{book-adzhemyan}. One
of these studies~\cite{basu04} has suggested, that multiscaling exponents might
depend on the cross correlations between the random forces in 3DRFMHD.

Here we wish to highlight, and examine in detail, the
implications of our study for intermittency. Some earlier DNS
studies, such as Refs.~\cite{mininni09,basu98}, had noted that,
for the case ${\rm Pr_M}=1$, the magnetic field is more
intermittent than the velocity field. This is why we have
concentrated on velocity and magnetic-field structure functions.
Our 3DRFMHD study confirms this finding, for the case ${\rm
Pr_M}=1$, as can be seen clearly from the comparison of our
exponent ratios, for ${\rm Pr_M}=1$, with those of the recent DNS
studies of 3DMHD turbulence~\cite{mininni09,sahoomhd,sahoophd}.  The error
bars that we quote for our exponent ratios have been calculated
as described in the previous Section.  Thus, at least given our
error bars, there is agreement between our 3DRFMHD exponent
ratios with those of 3DMHD~\cite{sahoomhd,sahoophd}. However, these error
bars are large, so a decisive comparison of multiscaling exponent
ratios of 3DRFMHD and 3DMHD turbulence must await
higher-resolution DNS studies than those presented here.
The ${\rm Pr_M}$ dependence of these exponent ratios 
is also similar for 3DRFMHD and 3DMHD turbulence.

Recent experimental studies of MHD turbulence in the solar
wind~\cite{salem09,podesta09} provide evidence for velocity
fields that are more strongly intermittent than the magnetic
field; this study does not give the value of ${\rm Pr_M}$.
However, their data for multiscaling exponents are qualitatively
similar to those we obtain at low values of ${\rm Pr_M}$. Of
course, we must exercise caution in comparing results from DNS
studies of homogeneous, isotropic, incompressible MHD turbulence
with measurements on the solar wind in which anisotropy, 
compressibility, and kinetic effects can be significant.

Moreover, we address is the issue of universality of
exponent ratios in 3DMHD and 3DRFMHD turbulence. This is of
central importance in deciding whether we can defensibly use the
latter as a model for obtaining universal statistical properties,
such as exponent ratios, by employing field-theoretic
renormalization-group methods. At the level of the DNS study we
have presented here for 3DRFMHD, these exponent ratios for
3DRFMHD and 3DMHD seem to agree, given our large error bars, in
all except a few cases (e.g., order $p=1$ and ${\rm
Pr_M}=1$)~\cite{sahoomhd,sahoophd} even though some PDFs are different in
detail and isosurfaces of $\omega$, $j$, etc., are qualitatively
different. Of course, as we have mentioned above, strictly
speaking 3DMHD and 3DRFMHD turbulence cannot be in the same
universality class because of logarithmic corrections to, say,
$E(k)$ that arise from the power law in the random forcing we
employ. Such logarithmic corrections have been discussed in the
analogous randomly forced versions of the 3D Navier-Stokes
equation~\cite{sain98} and the 1D Burgers equation~\cite{mitra05}.

\section*{Acknowledgements}
GS thanks AtMath Collaboration at the University of Helsinki.  We thank  D.
Mitra, S.S. Ray and A.K. Verma for discussions, SERC(IISc) for computational
resources and DST, UGC and CSIR India for support. 

Author contributions: GS, AB, and RP planned the study; GS and NBP carried out
the numerical simulations; NBP, GS, and RP analyzed the results; NBP, GS, and
RP wrote the manuscript; All authors checked and approved the final version of the
manuscript.

\end{document}